\documentclass[reprint,amsmath,amssymb,pre]{revtex4-2}

\usepackage{graphicx}
\usepackage{svg}

\usepackage{color}
\usepackage{dcolumn}
\usepackage{bm}
\usepackage{nicefrac}
\usepackage{hyperref}
\usepackage[textsize=tiny]{todonotes}

\renewcommand{\eqref}[1]{Eq.~(\ref{#1})}
\newcommand{\figref}[1]{Fig.~\ref{#1}}
\newcommand{\secref}[1]{Sec.~\ref{#1}}

\begin{document}

\title{Emergence of multiphase condensates from a limited set of chemical building blocks}

\author{Fan Chen}
\affiliation{Department of Chemistry, Princeton University, Princeton, NJ 08544, USA}
\author{William M.~Jacobs}
\email{wjacobs@princeton.edu}
\affiliation{Department of Chemistry, Princeton University, Princeton, NJ 08544, USA}

\date{\today}

\begin{abstract}
  Biomolecules composed of a limited set of chemical building blocks can co-localize into distinct, spatially segregated compartments known as biomolecular condensates.
  While many condensates are known to form spontaneously via phase separation, it has been unclear how immiscible condensates with precisely controlled molecular compositions assemble from a small number of chemical building blocks.
  We address this question by establishing a connection between the specificity of biomolecular interactions and the thermodynamic stability of coexisting condensates.
  By computing the minimum interaction specificity required to assemble condensates with target molecular compositions, we show how to design heteropolymer mixtures that produce compositionally complex condensates using only a small number of monomer types.
  Our results provide insight into how compositional specificity arises in naturally occurring multicomponent condensates and demonstrate a rational algorithm for engineering complex artificial condensates from simple chemical building blocks.
\end{abstract}

\maketitle

\section{Introduction}

Biomolecules such as proteins and nucleic acids can spontaneously co-localize to form multicomponent condensates via the thermodynamically driven process of phase separation~\cite{hyman2014liquid,shin2017liquid,berry2018physical,alberti2019considerations}.
A wide variety of different condensates have been observed in living cells, each of which is associated with a distinct macromolecular composition~\cite{banani2017biomolecular,xing2020quantitative}.
This \textit{compositional specificity} is central to the ability of condensates to selectively recruit client molecules~\cite{ditlev2018who,kilgore2022learning}, including enzymes and metabolites, and to respond dynamically to changes in macromolecular concentrations and the chemical state of the intracellular environment~\cite{su2016phase,case2019regulation,lafontaine2020nucleolus,sanders2020competing,wei2020nucleated,lyons2023functional}.
Compositional specificity is thus a key requirement for condensation to serve as a functional mechanism of self-organization in cellular biology.

Nonetheless, it has remained unclear how compositional specificity arises from conformationally disordered macromolecules~\cite{brangwynne2015polymer,harmon2017intrinsically,dignon2020biomolecular} and non-stoichiometric interactions in heterogeneous intracellular environments~\cite{pappu2023phase,musacchio2022role,villegas2022molecular}.
On the one hand, established principles of polymer physics predict that subtle differences in the chemical compositions of heteropolymers can drive demixing into coexisting phases~\cite{colby2003polymer}.
For example, random copolymers---i.e., heteropolymers composed of two monomer types---can demix into a large number of immiscible phases in accordance with their chemical compositions~\cite{nesarikar1993phase}.
Yet on the other hand, the molecular compositions of naturally occurring condensates appear to be considerably more complex~\cite{xing2020quantitative,sanders2020competing}, and the relationships between biomolecules and the condensates into which they partition is not always one-to-one~\cite{sanders2020competing,langdon2018mrna,ning2020drllps}.
These observations also raise the question of how well-studied driving forces for phase separation in model systems, such as arginine/tyrosine compositions of FUS condensates~\cite{wang2018grammar}, the aromatic composition of prion-like domain condensates~\cite{martin2020valence}, and many other sequence motifs~\cite{quiroz2015sequence,schuster2020identifying,greig2020arginine,krainer2021reentrant,ruff2022grammar,blazquez2023location,rekhi2024expanding}, coordinate condensate formation in heterogeneous mixtures, where a wide variety of biomolecules exhibit similar chemical features.
Addressing these questions is essential to understanding how biomolecular condensation can give rise to complex, tunable, and biologically functional spatial organization~\cite{jacobs2023theory}.

In this article, we focus on determining the minimum biomolecular interaction specificity that is required to achieve phase separation into non-trivial condensed phases in a multicomponent system.
In particular, we devise a theory to quantify and design ``minimal-complexity'' biomolecular mixtures---i.e., mixtures of heteropolymers built from the smallest number of distinct chemical building blocks and/or sequence motifs---that phase-separate into coexisting condensates with prescribed molecular compositions.
This approach builds on recent theoretical progress in modeling multicomponent phase separation~\cite{jacobs2023theory,jacobs2017phase,jacobs2021self,chen2023solver,harmon2018differential,mao2019phase,espinosa2020liquid,shrinivas2021phase,zwicker2022evolved,graf2022thermodynamic,carugno2022instabilities,thewes2023composition}, which aims to predict and engineer multiphase coexistence using simplified descriptions of biomolecular interactions.
The present work goes further by showing how compositional specificity in multicomponent mixtures can be achieved using a limited set of chemical building blocks.

\begin{figure*}
  \includegraphics[width=\textwidth]{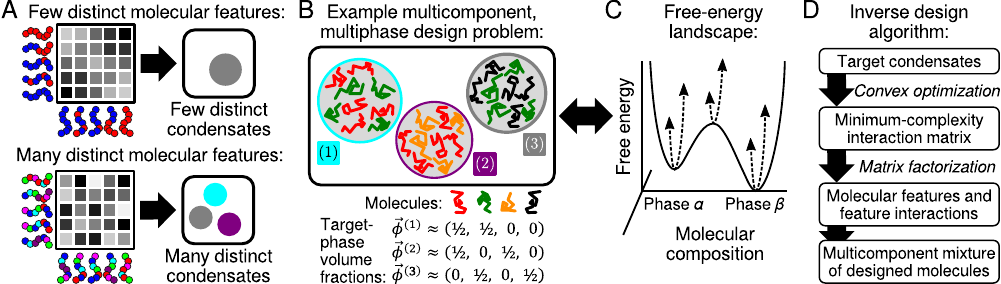}
  \caption{\textbf{Biomolecular interaction interdependencies and inverse design approach.}
    (A)~The interactions between biomolecules in a multicomponent mixture can be described by a pairwise interaction matrix.  In a system with few molecular features, such as polymers with a small number of distinct monomer types, the elements of the interaction matrix tend to be interdependent (top).  By contrast, in a mixture with many distinct molecular features, the interaction-matrix elements can be independently tuned (bottom).
    (B)~An example multiphase design problem.  Here, four molecular species (red, green, gold, and black) phase-separate into three immiscible condensates with prescribed molecular volume fractions $\{\vec\phi^{(\alpha)}\}$.
    (C)~A multiphase design problem corresponds to a high-dimensional free-energy landscape, where each target phase (such as the two shown here) specifies a local minimum.
    (D)~To solve the design problem, we first use convex optimization to find the \textit{minimum-complexity} interaction matrix, requiring the smallest number of distinct molecular features, that produces the desired free-energy landscape.  We then factorize the minimum-complexity interaction matrix to construct a multicomponent mixture of designed molecules.
    \label{fig:1}}
\end{figure*}

To this end, we first show how a description of biomolecular interaction specificity relates to the thermodynamic stability of coexisting multicomponent condensates.
We then introduce an ``inverse design'' approach~\cite{jacobs2021self,chen2023solver} for computing the minimum interaction specificity required to form immiscible condensates with prescribed molecular compositions.
We demonstrate how this design algorithm can be used to optimize mixtures of heteropolymers, subject to system-specific physicochemical constraints, that phase-separate into non-trivial coexisting phases while utilizing a surprisingly small number of distinct monomer types.
Finally, we perform molecular dynamics simulations of phase coexistence using these optimized heteropolymer mixtures to validate our theoretical predictions and inverse design approach.
Our results suggest an extensible and generalizable framework for exploring molecular ``grammars'' in multicomponent biomolecular systems~\cite{quiroz2015sequence,wang2018grammar,martin2020valence,schuster2020identifying,greig2020arginine,krainer2021reentrant,ruff2022grammar,blazquez2023location,rekhi2024expanding}, as well as a practical strategy for engineering complex artificial condensates using experimentally realizable molecules~\cite{simon2017programming,boeynaems2019spontaneous,fisher2020tunable,kaur2021sequence,leshem2023peptides,walls2023modular,dai2023synthetic,lu2020multiphase,do2022engineering,liu2023multiphasic}.

\section{Results}

\subsection{Describing interdependencies among biomolecular interactions}
\label{sec:interdependencies}

To describe biomolecular interaction specificity, we adopt a coarse-grained representation of biomolecules in a multicomponent fluid.
Specifically, we assume that \textit{pairwise} interactions can be used to describe the net attraction or repulsion between molecular species~\cite{jacobs2023theory}.
This approach implies that the chemical potential $\mu_i$ of each macromolecular species $i=1,\ldots,N$ depends linearly on the symmetric pairwise interaction matrix $\bm\epsilon$,
\begin{equation}
    \label{eq:mu}
    \mu_i(\vec{\phi}) = \mu_{\text{id},i}(\vec{\phi}) + \mu_{\text{v},i}(\vec{\phi}) + \sum_{j=1}^N \epsilon_{ij} \phi_j,
\end{equation}
where $\phi_i$ represents the molecular volume fraction (i.e., the concentration times the molecular volume, $v_i$) of each molecular species $i$.
The first two terms of \eqref{eq:mu} account for the ideal, $\mu_{\text{id},i} = v_i^{-1} k_{\text{B}}T \log\phi_i$, and steric, $\mu_{\text{v},i}$, contributions to the chemical potential, which are both independent of $\bm\epsilon$ (see SI).
We make the key approximation that $\bm\epsilon$ is independent of $\vec\phi$, meaning that the net interaction $\epsilon_{ij}$ between molecules of types $i$ and $j$ does not depend on the presence of other components.
This approximation is consistent with field-theoretic treatments of heteropolymer interactions~\cite{lin2018theories,wessen2022analytical}, sticker--spacer and patchy-particle models~\cite{pappu2023phase,jacobs2023theory}, and perturbative treatments of fluids~\cite{hansen2013theory} in the limit of weak monomer/sticker interactions.
Later on, we will use simulations to confirm that this approximation can make accurate predictions for multicomponent heteropolymer solutions.

The pairwise interaction matrix provides a systematic way of describing the interdependencies among biomolecular interactions in a multicomponent mixture.
In a full-rank interaction matrix, there are $N(N+1)/2$ independent pair interactions.
However, because biomolecular interactions are typically governed by a limited set of physicochemical features, it is conceivable that not all pair interactions are independent of one another.
For example, in a mixture of molecules that interact via hydrophobic forces alone, there might be only $N$ independent parameters---namely, the hydrophobicity of each molecular species---that determine the interaction matrix.
In general, we expect that various modes of interaction, including electrostatic, cation-$\pi$, and hydrogen bonding, among others~\cite{brangwynne2015polymer}, contribute to the net interactions among biomolecules in a multicomponent mixture.

Within the pairwise approximation, accounting for interaction interdependency (\figref{fig:1}A) implies factorizing $\bm\epsilon$ in terms of a molecular feature matrix, $W\in \mathcal{R}^{N\times r}$, and a feature interaction matrix, $\bm{u} \in \mathcal{R}^{r\times r}$~\cite{graf2022thermodynamic},
\begin{equation}
  \label{eq:WuW}
  \bm{\epsilon} = \bm{W} \bm{u} \bm{W}^\top.
\end{equation}
Here, a ``feature'' refers to a linearly independent molecular property, which could represent a literal molecular building block such as a nucleotide or an amino acid, or alternatively an emergent property such as a sequence motif or pattern involving adjacent monomers.
Each row vector of $\bm{W}$ describes the appearance of these features in a particular molecular species, while the matrix $\bm{u}$ describes the interactions among all pairs of features.
In other words, interactions between molecules arise due to the interactions, $\bm{u}$, among the molecular features, which form a basis set.
Molecular features are then combined into molecules via the $\bm{W}$ matrix, which embeds each molecule into the $r$-dimensional space of molecular features.
Importantly, the rank $r$ indicates the number of distinct features that are available, and thus controls the number of elements of $\bm\epsilon$ that can be independently tuned.

\subsection{Relating interaction interdependencies and biomolecular condensate thermodynamics}
\label{sec:landscape}

Next, to relate the number of available molecular features to the formation of coexisting biomolecular condensates, we consider a set of $K$ target condensed phases (i.e., condensates) with defined molecular compositions (\figref{fig:1}B).
Each target condensate $\alpha = 1,\ldots,K$ is described by a vector of molecular volume fractions $\vec\phi^{(\alpha)}$.
These condensed phases are in coexistence with a dilute phase, referred to as phase 0, with molecular volume fraction $\phi^{(0)}$.
For the condensed phases to be thermodynamically stable, each $\vec\phi^{(\alpha)}$ must correspond to a local minimum on a free-energy landscape $f(\vec\phi) - \vec\mu^{(0)}\cdot\vec\phi$, where $\mu_i^{(0)}$ is the chemical potential of molecular species $i$ in the dilute phase~\cite{hansen2013theory}.
From the relationship between the free-energy density $f$ and the chemical potentials, $\mu_i = \partial f / \partial\phi_i$, this stability condition implies that $\mu_i^{(\alpha)} = \mu_i^{(0)}$ for all condensed phases and molecular species.
The stability condition also implies that the Hessian of the free-energy density must be positive definite in each condensed phase $\alpha$, $\partial^2 f / \partial\phi_i \partial\phi_j|_{\vec\phi^{(\alpha)}} \succ 0$.

According to the free-energy landscape (\figref{fig:1}C), any mixture with overall macromolecular concentrations within the convex hull of the dilute-phase  and target-phase concentrations $\{\vec\phi^{(0)},\vec\phi^{(1)},\ldots,\vec\phi^{(K)}\}$ can phase-separate to form the prescribed condensed phases~\cite{chen2023solver}.
(Whether some or all of these phases form in practice depends on kinetic considerations, such as nucleation dynamics~\cite{jacobs2021self,jacobs2023theory}, that are not the focus of the present work.)
However, the pairwise interaction matrix that results in the prescribed condensed-phase molecular compositions is typically not unique.
In particular, there may be low-rank interaction matrices, for which $r < N$, that result in---or at least closely approximate---the target condensates.
We are interested in these low-rank interaction matrices, since they reduce the number of molecular features that are required to achieve the prescribed set of coexisting condensates within the pairwise approximation.

We start by considering a full-rank interaction matrix, $\bm\epsilon$, that results in the target set of condensed phases $\{\vec\phi^{(1)},\vec\phi^{(2)},\ldots,\vec\phi^{(K)}\}$.
Perturbing $\bm\epsilon$ by a matrix $\bm{\Delta\epsilon}$ results, to lowest order, in a shift of the condensed-phase compositions (see SI),
\begin{equation}
  \label{eq:perturbation}
  \Delta\vec{\phi}^{(\alpha)} = -\frac{1}{2} \bigg( \frac{\partial^2f}{\partial\phi_i\partial\phi_j} \bigg\rvert_{\vec{\phi}^{(\alpha)}} \bigg)^{\!-1} \!\bm{\Delta\epsilon}\, \vec{\phi}^{(\alpha)}.
\end{equation}
\eqref{eq:perturbation} says that the concentrations in the $\alpha$ phase are most sensitive to perturbations in the direction of the eigenvector, $\hat\nu_1^{(\alpha)}$, that corresponds to the minimum eigenvalue, $\lambda_1^{(\alpha)}$, of $\partial^2 f / \partial\phi_i \partial\phi_j|_{\vec\phi^{(\alpha)}}$.
We can therefore relate the Frobenius norm of a random matrix $\bm{\Delta\epsilon}$ to the relative change in the molecular compositions of the $\alpha$ phase in this least stable direction, $\eta \equiv 2 \Delta\vec\phi \cdot \hat\nu_1^{(\alpha)} / \|\vec\phi^{(\alpha)}\|$ (see SI),
\begin{equation}
  \label{eq:Deltaeps-bound}
  \|\bm{\Delta\epsilon}\|_F \approx \eta\lambda_1^{(\alpha)}.
\end{equation}
\eqref{eq:Deltaeps-bound} is useful because it relates the interaction-matrix perturbation to its effect on the phase behavior in the worst-case scenario.
Thus, to maintain the target molecular compositions to within a tolerance of $\eta$, we must limit deviations of the interaction matrix such that $\|\bm{\Delta\epsilon}\|_F \lesssim \min_\alpha \eta\lambda_1^{(\alpha)}$.

We can now determine the minimum required interaction specificity by attempting to reduce the number of molecular features, $r \le N$, used to represent the interaction matrix $\bm\epsilon$.
Doing so yields a rank-$r$ approximation, $\bm{\epsilon_r}$, for the $N \times N$ interaction matrix.
Here we apply the Eckart--Young--Mirsky (EYM) theorem~\cite{eckart1936rank}, which says that $\|\bm{\Delta\epsilon}\|_F \equiv \|\bm{\epsilon_r} - \bm{\epsilon}\|_F$ is minimized by eliminating the $N-r$ smallest singular values of $\bm\epsilon$.
We therefore find that the number of molecular features, $r$, must satisfy
\begin{equation}
  \label{eq:singval-bound}
  \left[\textstyle \sum_{k=1}^{N-r} \sigma_k^2 \right]^{1/2} \lesssim \min_\alpha \eta\lambda_1^{(\alpha)},
\end{equation}
where $\sigma_1 \le \sigma_2 \le \ldots \le \sigma_N$ are the singular values of $\bm\epsilon$, in order  to maintain the target molecular compositions.
In essence, the smallest $N-r$ singular values correspond to dimensions of the molecular feature space that are unnecessary for maintaining the target molecular compositions.

\eqref{eq:singval-bound}, which relates the singular values of the interaction matrix $\bm\epsilon$ to thermodynamic properties of coexisting condensates with prescribed molecular compositions, is the central result of our theoretical approach.
Importantly, a \textit{minimal complexity} mixture is one with the smallest number of molecular features, $r$, that satisfies \eqref{eq:singval-bound}.
\eqref{eq:singval-bound} therefore establishes a direct connection between the interdependencies of pairwise biomolecular interactions and the complexity of the phase-separated condensates that they can form, irrespective of specific details of the molecular features in a particular biomolecular system.

\subsection{Designing phase-separating heteropolymers using a minimal number of monomer types}
\label{sec:heteropolymers}

\begin{figure*} 
\centering
  \includegraphics[width=\textwidth]{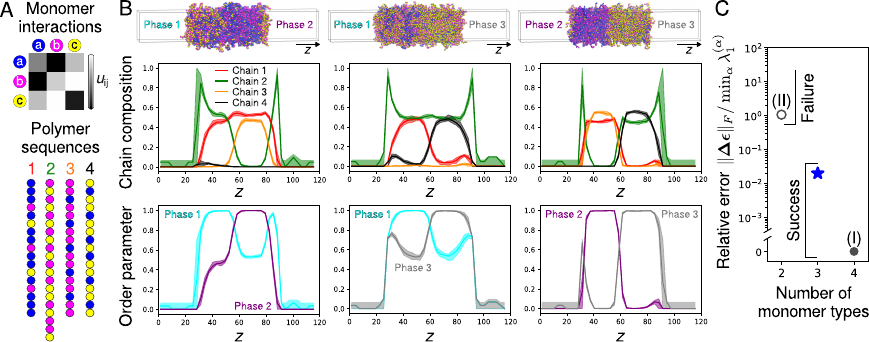}
  \caption{\textbf{Validation of minimal-complexity heteropolymer designs via molecular simulation.}
    (A)~A heteropolymer mixture designed to form coexisting condensates with the molecular compositions shown in \figref{fig:1}B.  According to \eqref{eq:singval-bound}, at most $r=3$ monomer types (a, b, and c) are required.
    (B)~Direct-coexistence simulations of all pairs of condensed phases in \figref{fig:1}B using the 3-monomer-type heteropolymer design shown in \textbf{A}.
    \textit{Top:} Equilibrium configurations, color-coded by monomer type.
    \textit{Middle:} Trajectory-averaged chain composition profiles along the direction of the simulation box perpendicular to the condensate interfaces.
    \textit{Bottom:} Corresponding trajectory-averaged order parameter profiles, \eqref{eq:order-parameter}.
    Shaded regions indicate the statistical uncertainty associated with the mean profiles.
    (C)~The relative reconstruction error (see text) for the heteropolymer mixture shown in \textbf{A--B} (blue star) and two alternative designs: (I) a successful (filled circle) 4-monomer-type design chosen to maximize the condensed-phase stabilities and (II) an unsuccessful (open circle) two-monomer-type design.  (See SI for complete details of these alternative designs.)
    \label{fig:3}}
\end{figure*}

\eqref{eq:singval-bound} implies that the target condensate compositions can be achieved---using the smallest possible number of distinct molecular features, $r$---by optimizing $\bm\epsilon$ to simultaneously minimize its $N-r$ smallest singular values and maximize the stabilities of the target condensed phases.
We can therefore use \eqref{eq:singval-bound} to solve an inverse design problem to find minimal-complexity biomolecular mixtures that phase-separate into condensates with non-trivial molecular composition (\figref{fig:1}D).
In practice, we first optimize a low-rank interaction matrix using convex optimization~\cite{jacobs2021self,chen2023solver} and then design a biomolecular mixture by factorizing $\bm\epsilon$, subject to appropriate physicochemical constraints.
(See SI for complete details.)

To demonstrate the predictive ability of our theory and its utility for designing complex phase-separated condensates, we apply this inverse design algorithm to a common simulation model of heteropolymers~\cite{kremer1990polymer}.
In this model, polymer species $i = 1, \ldots, N$ are linear chains of $L_i$ monomers, which are chosen from a library of $r$ monomer types (see SI for model details).
Nonbonded monomers of types $a,b = 1, \ldots, r$ interact via a cut-and-shifted Lennard-Jones (LJ) pair potential~\cite{lennardjones1924potential} with monomer diameter $d$.
Bonded monomers interact via finite-extensible nonlinear elastic (FENE) bonds~\cite{kremer1990polymer}.
In this context, minimal complexity designs are heteropolymer mixtures that require the fewest distinct monomer types.
Our aim is therefore to design mixtures of polymer sequences using the smallest number of monomer types, along with a nonpositive monomer interaction matrix $\bm{u^{\text{LJ}}}$ representing the attractive portion of the LJ pair potentials, in order to form the $K$ target condensates.

As an example, we first consider the 4-component, 3-condensed-phase mixture depicted in \figref{fig:1}B, which is non-trivial due to the ``enrichment'' (i.e., high target volume fractions) of chains A and B in two distinct, immiscible condensed phases.
Following the algorithm in \figref{fig:1}D (see SI), we apply convex optimization to find that at most $r=3$ distinct monomer types are required to stabilize condensates with these particular molecular compositions.
We then factorize an optimized rank-3 interaction matrix to find the $r \times r $ monomer interaction matrix $\bm{u^{\text{LJ}}}$ and the $N \times r$ feature matrix $\bm{W}$, which represents a count encoding of the number of occurrences of each monomer type within each heteropolymer sequence.
All elements of $\bm{W}$ are nonnegative integers, while the row sums of $\bm{W}$ are bounded by a maximum degree of polymerization $L_{\text{max}} = 20$.
Finally, we use the monomer interaction matrix to obtain the LJ pair potentials (with all attractive interactions weaker than the thermal energy), and we compose the polymer sequences from the count encoding matrix by interleaving the different monomer types to minimize the blockiness of each sequence.
An example outcome of this algorithm is shown in \figref{fig:3}A.

\subsection{Validating multicomponent, multiphase heteropolymer designs via molecular simulation}
\label{sec:simulation}

We test our heteropolymer mixture designs by performing direct-coexistence molecular dynamics simulations to examine the molecular compositions and immiscibility of the target condensates.
In these simulations, all chain types are present in a constant-temperature ($k_{\text{B}}T = 1$), constant-volume simulation with periodic boundary conditions (see SI).
However, the initial condition is chosen such that two different condensed phases, $\alpha \ne \beta$, are in contact with one another and also with a dilute phase.
The condensed phases are thermodynamically stable if they remain phase-separated and immiscible once the simulation has relaxed to equilibrium.
In practice, we ensure convergence by calculating simulation trajectories that are many times longer than the time required for all the chains in an $\alpha=\beta$ control simulation to mix completely (see SI).
We carry out simulations for all pairs of condensed phases $\alpha \ne \beta$ to confirm that they are mutually immiscible.
All calculations are performed using the LAMMPS simulation package~\cite{plimpton1995fast}, the Nose--Hoover thermostat, and a timestep of $5\times 10^{-3}$ in LJ units.

In \figref{fig:3}B, we show equilibrium configurations of simulations and polymer composition profiles for the LJ heteropolymer design shown in \figref{fig:3}A.
We define an order parameter for each phase $\alpha$ based on the cosine similarity of the local polymer composition $\vec\phi$ and the target $\vec\phi^{(\alpha)}$,
\begin{equation}
  \label{eq:order-parameter}
  q^{(\alpha)}(\vec\phi) \equiv \frac{ \vec\phi \cdot  \vec\phi^{(\alpha)} }{ ||\vec\phi|| \, ||\vec\phi^{(\alpha)}|| },
\end{equation}
to distinguish the coexisting condensed phases.
This order parameter is equal to one if the polymer composition of an equilibrated condensed phase matches that of target phase $\alpha$.
When coexisting condensed phases have ``shared'' components, meaning that one or more polymer species are enriched in multiple phases, the order parameters are not orthogonal, such that ${q^{(\alpha)}(\vec\phi^{(\beta)}) > 0}$.
The equilibrium order parameter profiles shown in \figref{fig:3}B confirm that this simulation model represents a solution to the design problem proposed in \figref{fig:1}B.

Importantly, our simulations indicate that the designed heteropolymer mixture shown in \figref{fig:3}A is not only a valid solution but also a minimal-complexity design.
Adding a fourth monomer type eliminates the reconstruction error, $\|\bm{\Delta\epsilon}\|_{\text{F}}$, that is introduced when eliminating one of the singular values in \eqref{eq:singval-bound} and factorizing the low-rank interaction matrix into polymer sequences with a discrete number of monomers of each type (\figref{fig:3}C).
Doing so provides additional control over the condensate interfaces, reducing the accumulation of chain type 2 at the condensate/dilute interfaces that occurs with the 3-monomer-type design (\figref{fig:3}B) and is not addressed by our theory, which focuses only on bulk phases.
However, this comes at the cost of increased heteropolymer complexity.
By contrast, using only two monomer types results in a large reconstruction error that violates \eqref{eq:singval-bound}, and, as predicted, the simulated condensed phases are observed to mix.
This observation suggests that $r=3$ is indeed the minimal-complexity requirement for realizing the coexisting condensates shown in \figref{fig:1}B.
(See SI for detailed descriptions of these alternative designs and accompanying simulation data.)

\subsection{Evaluating the pairwise approximation using simulations of successful condensate designs}
\label{sec:correlations}

\begin{figure}  
  \includegraphics[width=\columnwidth]{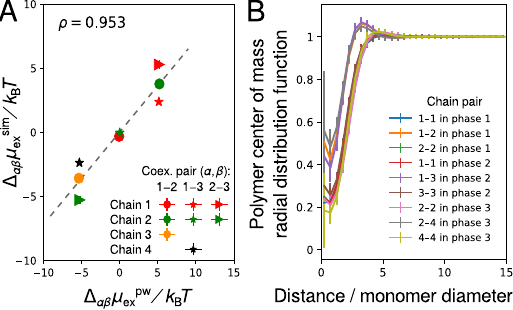}
  \caption{\textbf{Validation of the pairwise-additive approximation for polymer--polymer interactions.}
    (A)~Correlation between the excess chemical potential differences measured in direct-coexistence simulations (sim) and predicted by the pairwise approximation (pw).  Points are shown for chain types in coexisting condensed phase pairs ($\alpha,\beta$) along with the Pearson correlation coefficient, $\rho$.
    (B)~Radial distribution functions between the centers of mass of chains for all chain types that are enriched in each of the condensed phases.
    \label{fig:4}}
\end{figure}

Given this successful simulation test of a minimal-complexity heteropolymer mixture, we next assess accuracy of the pairwise-additive approximation in this system.
We first analyze the excess chemical potentials, $\mu_{\text{ex},i} \equiv \mu_i - \mu_{\text{id},i}$, of the heteropolymers predicted by \eqref{eq:mu}.
From our simulations, we extract the excess chemical potential differences, $\Delta_{\alpha \beta}\mu^{\text{sim}}_{\text{ex}, i}$, from the equilibrium compositions of simulated $\alpha$ and $\beta$ bulk phases and compare with the pairwise predictions, $\Delta_{\alpha \beta}\mu^{\text{pw}}_{\text{ex}, i}$ (see SI).
The resulting Pearson correlation coefficient of 0.93 (\figref{fig:4}A) indicates that the pairwise approximation is highly accurate in the condensed phases and follows from the interaction matrix, $\bm\epsilon$.
This observation is consistent with Flory's screening hypothesis~\cite{de1979scaling}, which likely applies to our finite-chain-length heteropolymer mixtures because the overlap parameter is reasonably large in the condensed phases ($P \approx 13 \pm 2$)~\cite{colby2003polymer}.
Thus, while the attractive contributions to the monomer--monomer pair potentials are much weaker than the thermal energy, each chain makes a large number of monomer--monomer contacts in the condensed phases.

\begin{figure*}
\centering
\includegraphics[width=\textwidth]{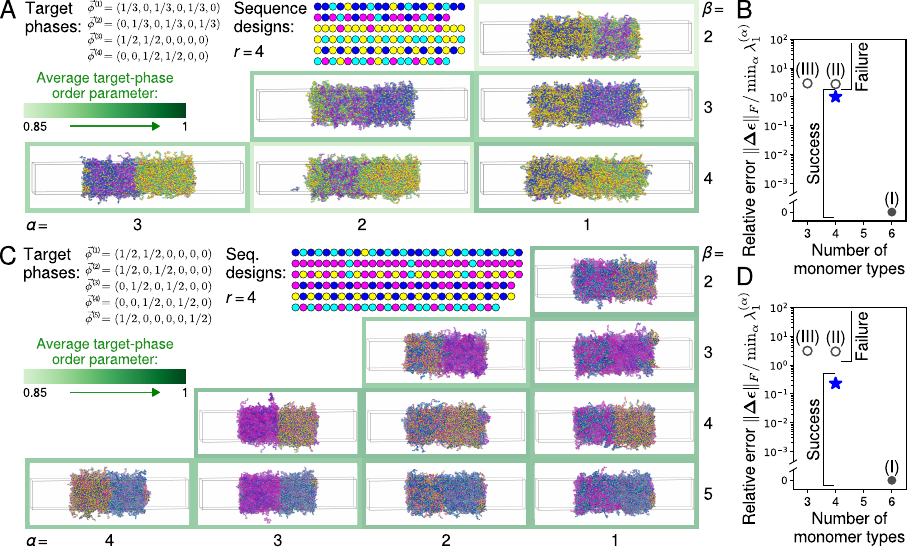}
\caption{\textbf{Design of complex multicomponent, multiphase condensates.}
  (A)~Inverse design and simulation validation of a 6-component, 4-condensed-phase system with the indicated target-phase molecular volume fractions.  Sequence designs use the minimal required number of monomer types, $r=4$.  Equilibrium configurations of $\alpha$--$\beta$ direct-coexistence simulations are shown as in \figref{fig:3}A, where the agreement between the order parameter profile and the target composition is indicated by the border color (see SI).
  (B)~The relative reconstruction error for the heteropolymer mixture shown in \textbf{A} (blue star) and three alternative designs: (I) a successful (filled circle) 4-monomer-type design chosen to maximize the condensed-phase stabilities and unsuccessful (open circle) designs with (II) four or (III) three monomer types, which do not satisfy \eqref{eq:singval-bound}.
  (C)~Inverse design and simulation validation of a 6-component, 5-condensed-phase system with the indicated target-phase molecular volume fractions.  Here, the minimal required number of monomer types is also $r=4$.
  (D)~The relative reconstruction error for the mixture shown in \textbf{C} (blue star) and alternative designs as in \textbf{B}.  Complete data for all designs are provided in the SI.
   \label{fig:5}}
\end{figure*}

We further find that the microstructures of the condensed phases agree with expectations based on the Gaussian Core Model (GCM) of finite-length polymers~\cite{louis2000gcm}.
From simulations of homogeneous condensed phases, we obtain the radial distribution function (RDF) for the center of mass of the polymer chains (see SI).
Consistent with the GCM, the RDFs for all polymer species pairs exhibit a correlation hole, meaning that the centers of mass overlap with low probability, while a positive peak is found at a distance of approximately twice the radius of gyration, implying effective attractions between chains in the condensed phases.
The amplitudes of these peaks anticorrelate with the predicted pairwise interaction strengths given by $\bm\epsilon$ (\figref{fig:4}B), while the zero-wavenumber structure factor, which can be directly related to the pairwise interactions~\cite{hansen2013theory}, exhibits a Pearson correlation coefficient of 0.996 with respect to the predictions of the pairwise approximation (see SI).
Taken together, our simulation results indicate that the mean-field approximations underlying our theory and design approach are sufficiently accurate to engineer chemically realistic heteropolymer mixtures with complex phase behavior.

\subsection{Designing minimal-complexity biomolecular mixtures with complex phase behavior}

To further demonstrate the capabilities of our approach, we apply our inverse design algorithm to two additional interesting scenarios and verify our predictions using simulations of LJ heteropolymers (see SI for complete details).
First, we consider a 6-component mixture in which components are shared among four condensates with a nonuniform number of enriched components per phase (\figref{fig:5}A).
Following the approach described in \secref{sec:heteropolymers}, we predict that at most $r=4$ monomer types are required to construct LJ heteropolymers that phase separate into the prescribed phases.
We then confirm this prediction by performing simulations of all pairs of condensed phases as described in \secref{sec:simulation}.
We also consider alternative designs with a variable number of monomer types in \figref{fig:5}B, including an alternative 4-monomer-type design whose factorized interaction matrix does not satisfy \eqref{eq:singval-bound}.
These results are consistent with the prediction that at most four monomer types are required and highlight the importance of following our design algorithm to obtain a successful minimal-complexity design.

Second, we examine a scenario in which a 6-component mixture forms five condensed phases with shared components (\figref{fig:5}C).
Here we predict that the required number of molecular features, $r=4$, is not only smaller than the number of components, but is also smaller than the number of condensed phases.
However, to find a factorized interaction matrix that satisfies \eqref{eq:singval-bound}, we find that it is necessary to increase the maximum degree of polymerization to $L_{\text{max}} = 30$.
Again our simulations of designed LJ heteropolymers (\figref{fig:5}C) and comparisons to alternative designs (\figref{fig:5}D) support our prediction that at most four monomer types are required to achieve the desired phase behavior.
These additional examples indicate that rationally designed low-complexity heteropolymer mixtures can indeed achieve highly non-trivial compositional specificity via phase separation.

\section{Discussion}
\label{sec:discussion}

The central result of this work is a relationship between the compositional specificity of phase-separated condensates and the complexity of the pairwise interactions in a biomolecular mixture.
Achieving complex self-organization via phase separation requires a minimal number of distinct molecular features, which represent linearly independent modes of interaction in a biomolecular mixture.
We have shown that for systems in which the pairwise approximation is sufficiently accurate, this required number of molecular features can be determined directly from the target phase behavior.
We have also developed a practical inverse design algorithm for calculating an upper bound on the required number of molecular features and then constructing minimal-complexity biomolecular mixtures that phase-separate into condensates with the target molecular compositions.
Extensive simulations of sequence-dependent heteropolymers support the predictions of our theory and demonstrate the effectiveness of our inverse design approach.

We emphasize that this approach is useful because it can be applied to understand and rationally design coexisting condensates \textit{with non-trivial molecular compositions}, including scenarios in which a single molecular component is enriched in multiple phases.
This feature is crucial in the context of intracellular biomolecular condensates, whose intricate molecular compositions~\cite{banani2017biomolecular,xing2020quantitative,ditlev2018who,kilgore2022learning} do not appear to resemble the phase behavior of simpler heteropolymer mixtures, such as random copolymers~\cite{nesarikar1993phase}.
Nonetheless, similarly to random copolymers, we find that a relatively small number of monomer types is typically required to generate many coexisting phases even when non-trivial molecular compositions are prescribed.
Consequently, constructing a \textit{minimal-complexity} heteropolymer mixture, which utilizes the predicted minimum number of distinct monomer types while maintaining compositional specificity of the coexisting condensates, can substantially reduce the effective dimension of the pairwise interaction matrix.

We anticipate that extensions of our approach will provide a unifying conceptual framework for unravelling the ``molecular grammars'' governing phase separation in naturally occurring biomolecular mixtures~\cite{brangwynne2015polymer,quiroz2015sequence,wang2018grammar,martin2020valence,schuster2020identifying,greig2020arginine,krainer2021reentrant,ruff2022grammar,blazquez2023location,rekhi2024expanding}.
It is important to note that the minimal number of molecular features predicted by our theory provides an upper bound on the required number of monomer types, since motifs composed of adjacent monomers can potentially give rise to additional independent molecular features within the pairwise approximation.
Moreover, significant contributions from non-pairwise interactions, particularly in nucleic acid~\cite{jain2017rna} and multi-domain protein~\cite{li2012phase} mixtures, may allow for compositional specificity with even fewer distinct molecular features.
For example, it is well known that sequence-dependent properties, such as sequence ``blockiness''~\cite{martin2020valence,rana2021blocky} and charge-patterning effects~\cite{das2013conformations,lin2016sequence,pal2021patterning,chew2023thermodynamic}, tune the phase behavior of heteropolymer solutions in ways that cannot be predicted by the monomer compositions of the heteropolymers alone.
Non-pairwise interactions also arise due to secondary-structure formation~\cite{conicella2016mutations,hegde2023competition} and strong one-to-one associative binding interactions~\cite{semenov1998thermoreversible}.
In future work, we will incorporate such sequence-dependent features into our inverse design approach in order to tighten the bound on the required number of monomer types and broaden the range of applicability of our heteropolymer design algorithm.

In conclusion, our theory and inverse design approach advance our understanding of multiphase condensate formation in biology, as well as our ability to engineer chemically diverse artificial condensates.
Our predictions could be tested using existing experimental technologies in both biological~\cite{simon2017programming,boeynaems2019spontaneous,fisher2020tunable,kaur2021sequence,leshem2023peptides,walls2023modular} and synthetic~\cite{dai2023synthetic,lu2020multiphase,do2022engineering,liu2023multiphasic} systems.
Our theory could also be applied to design ``patchy'' colloidal particles~\cite{glotzer2007anisotropy}, in which case the feature matrix would indicate the surface area covered by a sticker type and the feature--feature interactions would represent the interactions between different types of stickers.
Ultimately, by incorporating the design of tunable interfacial tensions~\cite{mao2020designing,li2023predicting} and nonequilibrium chemical activity~\cite{cho2023tuning,cho2023nonequilibrium} into this framework, it will be possible to design fully ``programmable'' fluids and soft materials that self-organize via multicomponent phase separation.

\begin{acknowledgments}
  The authors acknowledge Ofer Kimchi, Isabella Graf, and Zhuang Liu for providing comments on a previous version of this manuscript and Princeton Research Computing for technical support.
  Source code and example calculations can be found at \texttt{https://github.com/wmjac/phaseprogramming-pub}.
\end{acknowledgments}

\providecommand{\latin}[1]{#1}
\makeatletter
\providecommand{\doi}
  {\begingroup\let\do\@makeother\dospecials
  \catcode`\{=1 \catcode`\}=2 \doi@aux}
\providecommand{\doi@aux}[1]{\endgroup\texttt{#1}}
\makeatother
\providecommand*\mcitethebibliography{\thebibliography}
\csname @ifundefined\endcsname{endmcitethebibliography}
  {\let\endmcitethebibliography\endthebibliography}{}

\end{document}


\maketitle

\onecolumngrid
\renewcommand\thefigure{S\arabic{figure}}    
\setcounter{figure}{0}
\renewcommand\theequation{S\arabic{equation}}    
\setcounter{equation}{0}

\section*{Supplementary Information for\\``Emergence of multiphase condensates from a limited set of chemical building blocks''}

\section{Relating interaction interdependencies and biomolecular condensate thermodynamics}

\subsection{Multicomponent macromolecular solutions with pairwise interactions}
\label{sec:pairwise}

Throughout this work, we consider a solution of $N$ macromolecular species in an implicit solvent.
The concentrations (i.e., number densities) of the macromolecular species, $\{\rho_i\}$, are related to the macromolecular volume fractions, $\{\phi_i \equiv v_i \rho_i\}$, by the macromolecular excluded volumes, $\{v_i\}$.
As discussed in the main text, we assume a free-energy density $f$ in which the molecular interactions follow a pairwise approximation.
This means that the (scaled) chemical potentials of all macromolecular components, $\mu_i = \partial f / \partial \phi_i = v_i^{-1} \partial f / \partial \rho_i$, can be described by the equation
\begin{equation}
    \label{eq:mu}
    \mu_i(\vec{\phi}) = \mu_{\text{id},i}(\vec{\phi}) + \mu_{\text{v},i}(\vec{\phi}) + \sum_{j=1}^N \epsilon_{ij} \phi_j,
\end{equation}
where $\mu_{\text{id},i}(\vec\phi) = v_i^{-1} k_{\text{B}}T \log\phi_i$ is the ideal contribution to the chemical potential, $\mu_{\text{v},i}(\vec\phi)$ accounts for steric contributions, and the pairwise interaction matrix, $\bm\epsilon$, is assumed to be independent of the macromolecular concentrations.
More specifically, $\mu_{\text{v},i}(\vec\phi)$ represents the excess chemical potential due to short-ranged (e.g., hard-core) repulsive interactions between the macromolecules at finite concentration (and between the macromolecules and the solvent), while the pairwise interactions only describe longer ranged attractive or repulsive interactions.
This decomposition follows the highly successful perturbative description of liquids developed by Weeks, Chandler, and Anderson~\cite{weeks1971role}.
An explicit example will be given in SI~\secref{sec:heteropolymer}.
Note that the chemical potential of macromolecular species $i$, as defined above, is scaled with respect to the excluded volume $v_i$; this choice is made for convenience and does not affect our results.
Both the ideal and the steric contributions are independent of the pairwise interaction matrix, $\bm\epsilon$.

\subsection{Perturbative analysis of the landscape stability}
\label{sec:perturbative}

Applying small perturbations to the component-wise interactions $\{\epsilon_{ij}\}$, we expand the free-energy density, $f$, to linear order,
\begin{equation}
  \label{eq:f-expansion}
  \tilde{f} = f + \sum_{i=1}^N \sum_{j = i}^N \frac{\partial f}{\partial \epsilon_{ij}} \Delta \epsilon_{ij} + \sum_{i=1}^N \frac{\partial f}{\partial \phi_i} \Delta \phi_i.
\end{equation}
Since $\bm \epsilon$ is a symmetric $N \times N$ matrix, we only consider the independent degrees of freedom $\{\epsilon_{ij}\}$ for which $i \leq j$.  
\eqref{eq:mu} implies that $\partial f / \partial\epsilon_{ij} = (1/2)^{\delta_{ij}}\phi_i\phi_j$.
We can therefore rewrite the term in \eqref{eq:f-expansion} involving $\Delta \epsilon_{ij}$ in matrix--vector notation, 
\begin{equation}
  \label{eq:matrix}
  \begin{aligned}
    \sum_{i=1}^N \sum_{j = i}^N \frac{\partial f}{\partial \epsilon_{ij}} \Delta \epsilon_{ij} &= \frac{1}{2} \bigg(\sum_{i=1}^N \phi_i \Delta\epsilon_{ii}\phi_i + 2\sum_{i< j} \phi_i \Delta\epsilon_{ij} \phi_j  \bigg)\\
    & = \frac{1}{2} \bigg( \sum_{i=1}^N \phi_i \Delta\epsilon_{ii}\phi_i + \sum_{i< j} \phi_i \Delta\epsilon_{ij} \phi_j + \sum_{i > j} \phi_i \Delta\epsilon_{ij} \phi_j \bigg) \\
    & = \frac{1}{2}{\vec{\phi}}^\top\! \Delta \bm\epsilon\vec{\phi}.
  \end{aligned}
\end{equation} 
We then expand the chemical potentials at phase coexistence,
\begin{equation}
  \tilde{\mu}_k = \mu_k +
  \sum_{j=i}^N\sum_{i=1}^N \frac{\partial^2 f}{\partial \epsilon_{ij} \partial \phi_k} \Delta \epsilon_{ij} 
  + \sum_{i=1}^N \frac{\partial^2 f}{\partial\phi_k \partial\phi_i}\Delta\phi_i.
\end{equation}
The change in the chemical potentials, $\Delta\vec\mu$, can be written in matrix--vector notation as
\begin{equation}
  \Delta\vec{\mu} \equiv \tilde{\vec\mu} - \vec\mu = \Delta \bm{\epsilon} \vec{\phi} + \left(\frac{\partial^2 f}{\partial {\vec\phi}^{\,2}}\right) \Delta \vec{\phi},
  \label{eq:dmu}
\end{equation}
where the $\Delta\bm\epsilon\Vec{\phi}$ term is obtained by simplifying the expression
\begin{equation}
  \begin{aligned}
    \sum_{j=1}^N\sum_{i=1}^N \frac{\partial^2 f}{\partial \epsilon_{ij} \partial \phi_k} \Delta \epsilon_{ij} &=  
    \sum_{j=1}^N\sum_{i=1}^N (1/2)^{\delta_{ij}} (\delta_{ik}\phi_j + \delta_{jk}\phi_i) \epsilon_{ij}\\
    &= \sum_{j=1}^N (1/2)^{\delta_{kj}} \epsilon_{kj}\phi_j + \sum_{i=1}^N (1/2)^{\delta_{ik}} \epsilon_{ki}\phi_k  \\
    &= \frac{1}{2}\sum_{j\neq k} \Delta\epsilon_{kj}\phi_j + \frac{1}{2}\sum_{i\neq k} \Delta\epsilon_{ki}\phi_i + \Delta\epsilon_{kk}\phi_k \\
    &= (\bm{\Delta\epsilon} \vec{\phi})_k.
  \end{aligned}
  \label{eq:}
\end{equation}
The grand potential density in the $\alpha$ phase, $\tilde\Omega^{(\alpha)} \equiv \tilde f - \tilde{\vec\mu} \cdot \vec\phi$, can similarly be written as
\begin{equation}
  \label{eq:lse}
  \begin{aligned}
    \tilde\Omega^{(\alpha)} &= \Omega^{(\alpha)} + \left.\frac{\partial f}{\partial \vec \phi}\right|^{(\alpha)} \cdot \Delta \vec\phi^{(\alpha)} + \sum_{i=1}^N \sum_{j=i}^N \left.\frac{\partial f}{\partial \epsilon_{ij}}\right|^{(\alpha)}\!\!\Delta\epsilon_{ij} - \vec\mu \cdot \Delta\vec\phi^{(\alpha)} - \Delta\vec\mu^{(\alpha)} \cdot \vec\phi^{(\alpha)} \\
    &= \Omega^{(\alpha)} + \vec\mu \cdot \Delta\vec\phi^{(\alpha)} + \frac{1}{2} {\vec{\phi}^{(\alpha)}}^\top\! \Delta\bm\epsilon \vec{\phi}^{(\alpha)} - \vec\mu \cdot \Delta\vec\phi^{(\alpha)} - \Delta\vec\mu^{(\alpha)} \cdot \vec\phi^{(\alpha)} \\
    &= \Omega^{(\alpha)} + \frac{1}{2} {\vec{\phi}^{(\alpha)}}^\top\! \Delta\bm\epsilon \vec{\phi}^{(\alpha)} - \Delta \vec\mu^{(\alpha)} \cdot \vec\phi^{(\alpha)}.
\end{aligned}
\end{equation}
At coexistence, it is required that $\vec\mu^{(\alpha)} = \vec\mu^{(0)}$ and $\Omega^{(\alpha)} = \Omega^{(0)}$ for every condensed phase $\alpha = 1, \ldots, K$, where the index 0 indicates the dilute phase.
Requiring that the perturbed phases remain at coexistence means that $\Delta \vec{\mu}^{(\alpha)} = \Delta \vec{\mu}^{(0)}$ and $\tilde\Omega^{(\alpha)} = \tilde\Omega^{(0)}$.
Since $\Delta \vec{\mu} = \Delta \bm{\epsilon} \vec{\phi} + (\partial^2 f / \partial {\vec{\phi}}^2) \Delta \vec{\phi} $ from \eqref{eq:dmu}, we find that the coexistence condition for $\tilde\Omega$ is
\begin{equation}
  -\frac{1}{2}(\vec{\phi}^{(\alpha)} - \vec{\phi}^{(0)})^\top  \Delta\bm\epsilon (\vec{\phi}^{(\alpha)} - \vec{\phi}^{(0)}) = 
  (\vec{\phi}^{(\alpha)} - \vec{\phi}^{(0)})^\top \bigg(\frac{\partial^2 f}{\partial {\vec{\phi}}^{\,2}} \bigg\rvert^{(\alpha)} \bigg) \Delta \vec{\phi}^{(\alpha)}.
\end{equation}
Assuming that the dilute phase is very dilute, this result simplifies to
\begin{equation}
  \label{eq:response}
  \Delta \bm\epsilon \vec{\phi}^{(\alpha)} = - 2 \frac{\partial^2 f}{\partial {\vec{\phi}}^{\,2}} \bigg\rvert^{(\alpha)} \!\cdot \Delta\vec{\phi}^{(\alpha)}.
\end{equation}
\eqref{eq:response} relates changes in the macromolecular concentrations in the $\alpha$ phase to perturbations in the pairwise interactions, assuming that the Hessian matrix of the $\alpha$ phase is known.
We can then estimate the noise tolerance of the interaction matrix $\bm\epsilon$ in the worst-case scenario by considering compositional changes along the least stable direction, $\hat{\nu}_1^{(\alpha)}$, of each target phase, where
\begin{equation}
  \frac{\partial^2 f}{\partial {\vec{\phi}}^{\,2}} \bigg\rvert^{(\alpha)} \!\cdot \hat{\nu}_1^{(\alpha)} = \lambda_1^{(\alpha)} \hat{\nu}_1^{(\alpha)},
\end{equation}
and the eigenvalues of the Hessian matrix are $0 > \lambda_1^{(\alpha)} \ge  \lambda_2^{(\alpha)} \ge \ldots \ge  \lambda_N^{(\alpha)}$.
Projecting an arbitrary concentration change $\Delta\vec\phi$ onto the least stable direction, we find
\begin{equation}
\| \Delta\bm{\epsilon} \vec{\phi}^{(\alpha)}\|_F = \eta \lambda_1^{(\alpha)}, 
\end{equation}
where $\eta \equiv 2\Delta\vec\phi \cdot \hat{\nu}_1^{(\alpha)}/\|\vec\phi^{(\alpha)}\|$ corresponds to twice the relative percentage change of macromolecular composition in the $\alpha$ phase and $\|\cdot\|_F$ is the Frobenius norm.
Applying the sub-multiplicative property of the Frobenius norm to the left-hand side, we have
\begin{equation*}
\| \Delta\bm{\epsilon} \hat{\phi}\|_F \leq \| \Delta\bm{\epsilon}\|_F \|\hat{\phi}\|_2;
\end{equation*}
the equality holds if $\hat\phi$ and each row of $\Delta\bm\epsilon$ are linearly independent. 
Assuming that this is the case, we can express the tolerance of the reconstruction error for each target phase $\alpha$ in terms of an allowed relative composition change $\eta$, 
\begin{equation}
  \|\Delta\bm{\epsilon}\|_F \approx \eta \lambda_1^{(\alpha)}.
\end{equation}
The overall tolerance is therefore given by 
\begin{equation}
  \label{eq:tolerance}
  \|\Delta\bm{\epsilon}\|_F \approx \min_\alpha \eta \lambda_1^{(\alpha)}.
\end{equation}

The optimal interaction matrix $\bm\epsilon$ should simultaneously maximize the landscape stability and minimize the reconstruction error, $\|\Delta\bm{\epsilon}\|_F$.
Using \eqref{eq:tolerance}, we can apply the Eckart--Young--Mirsky (EYM) theorem~\cite{eckart1936rank} to find the smallest number of distinct molecular features, $r$, that successfully solve the inverse design problem,
\begin{equation}
  \label{eq:singval-bound}
  \left[\textstyle \sum_{k=1}^{N-r} \sigma_k^2 \right]^{1/2} \lesssim \min_\alpha \eta\lambda_1^{(\alpha)},
\end{equation}
where $\sigma_1 \le \sigma_2 \le \ldots \le \sigma_N$ are the singular values of $\bm\epsilon$.
In other words, for a given $N \times N$ interaction matrix $\bm\epsilon$, the rank-$r$ approximation, $\bm{\epsilon_r}$, that minimizes $\|\bm{\Delta\epsilon}\|_F = \|\bm{\epsilon_r} - \bm\epsilon\|_F$ can be obtained by eliminating the smallest $N-r$ singular values of $\bm\epsilon$.
If these smallest singular values are nonzero, then the left-hand-side of \eqref{eq:singval-bound} will be nonzero, and the rank-$r$ approximation of the interaction matrix, $\bm{\epsilon_r}$, will not exactly equal the original interaction matrix, $\bm\epsilon$.
However, the macromolecular compositions of the resulting coexisting phases will deviate from the target compositions within the allowed tolerance $\eta$ if \eqref{eq:singval-bound} is satisfied.
\eqref{eq:singval-bound} thus motivates the identification of the minimal number of distinct molecular features, $r$, needed to stabilize a set of target phases.

\section{Two-step optimization-based inverse design approach}

\subsection{Step 1: Optimization of component-wise interactions.}
\label{sec:step1}

\subsubsection{Overview of convex optimization for minimal-complexity interactions}

In order to obtain coexisting phases with prescribed macromolecular volume fractions $\{\vec\phi^{(1)}, \vec\phi^{(2)}, \ldots, \vec\phi^{(K)}\}$, we first solve for the component-wise interactions using convex programming.
This step introduces controlled approximations to transform the nonlinear thermodynamic stability and design constraints into a convex optimization problem.
This convex relaxation is defined by the target condensed-phase volume fractions $\{\vec\phi^{(1)},\vec\phi^{(2)},\ldots,\vec\phi^{(K)}\}$, the molecular volumes of the $N$ molecular species, and a minimum stability criterion $\lambda_{\text{min}}$, which imposes a lower bound on $\lambda_1^{(\alpha)}$ for each condensed phase $\alpha$.
By satisfying the constraints of this convex optimization problem, we identify a ``solution space'' of pairwise interaction matrices that closely approximate the target free-energy landscape.
Importantly, convexity implies that this solution space can be efficiently computed, or otherwise proven to be infeasible if solutions to the inverse design problem do not exist~\cite{boyd2004convex,diamond2016cvxpy,odonoghue2016conic}.
The convex relaxation is defined explicitly in the following section, and we refer the reader to Ref.~\cite{chen2023solver} for a detailed discussion of this approach.

Within this solution space of pairwise interaction matrices, we wish to find a matrix with the fewest required number of molecular features, in accordance with \eqref{eq:singval-bound}.
We therefore minimize the nuclear norm of the interaction matrix, $\|\bm\epsilon\|_* \equiv \sum_{k=1}^N\sigma_k$, which is a convex relaxation of the matrix rank~\cite{recht2010minrank}, subject to the aforementioned convex constraints.
Including this objective function tends to reduce the magnitudes of the smallest singular values of $\bm\epsilon$.
Moreover, minimizing $\|\bm\epsilon\|_*$ guarantees that the solution to the convex optimization problem is unique.
We can then determine an upper bound on the minimum required number of features, $r$, using \eqref{eq:singval-bound}, and finally construct a rank-$r$ approximation of the interaction matrix, $\bm{\epsilon_r}$, by applying the EYM theorem.

\subsubsection{Semi-definite program for component-wise interactions}

The convex relaxation of the thermodynamic-stability and target-volume-fraction constraints defines a semi-definite program (SDP)~\cite{boyd2004convex},
\begin{subequations}
  \label{eq:constraints}
  \begin{align}
    \label{eq:constraint-eqmu}
    \mu_{\text{id},i}(\vec\phi^{(\alpha)}; \vec{v}) + \mu_{\text{ex},i}(\vec\phi^{(\alpha)}; \bm{\epsilon}, \vec{v}) &\ge \mu_i \;\forall i,\alpha \\
    \label{eq:constraint-eqP}
    P(\vec\phi^{(\alpha)}; \bm\epsilon, \vec{v}) &= 0 \;\forall \alpha \\
    \label{eq:constraint-possdef}
     \partial [\vec\mu_{\text{id}}(\vec\phi^{(\alpha)}; \vec{v}) + \vec\mu_{\text{ex}}(\vec\phi^{(\alpha)}; \bm{\epsilon}, \vec{v})] / \partial \vec\phi &\succ \lambda_{\text{min}}I \;\forall \alpha \\
     \label{eq:constraint-crit}
     \phiT^{(0)}(\vec\mu;\vec v) &< \phiT^*(\vec v).
  \end{align}
\end{subequations}
Components are either ``enriched'' or ``depleted'' in the $\alpha$ phase, depending on whether they make a substantial or negligible contribution, respectively, to the macromolecular composition of that phase.
The ideal chemical potential is $\mu_{\text{id},i} = v_i^{-1}\log\phi_i^{(\alpha)}$ for any component $i$ that is enriched in the $\alpha$ phase or $\mu_{\text{id},i} = v_i^{-1}\log\phi_{\text{depl}}^{(\alpha)}$ for any component $i$ that is depleted in the $\alpha$ phase.
The equality(inequality) in \eqref{eq:constraint-eqmu} applies to enriched(depleted) components.
The volume fraction of any component that is depleted in the $\alpha$ phase cannot exceed $\phi_{\text{depl}}^{(\alpha)} \equiv \zeta\phiT^{(\alpha)}/[M^{(\alpha)}(N-M^{(\alpha)})]$, where $\phiT^{(\alpha)}$ is the total volume compositions of all components in the $\alpha$ phase, $M^{(\alpha)}$ is the number of enriched components in the $\alpha$ phase, and $\zeta$ is an adjustable parameter.
In this work, we choose $\phiT \ge 0.9$ and $\zeta = 10^{-2}$ for all target phases when performing this optimization step.
\eqref{eq:constraint-eqP} states that the pressure, $P$, must be zero in all phases, since the pressure in the dilute phase, which is nearly ideal, is also approximately zero.
In \eqref{eq:constraint-possdef}, the parameter $\lambda_{\text{min}} \ge 0$ places a lower bound on the smallest eigenvalue of the Hessian matrix in order to guarantee thermodynamic stability.
The final constraint, \eqref{eq:constraint-crit}, ensures that the volume fraction in the dilute phase, $\phiT^{(0)}$, is less than the critical volume fraction, $\phiT^*(\vec{v})$.
This condition is independent of $\bm\epsilon$ due to the zero-osmotic-pressure assumption.
We implement and solve this SDP in practice using efficient convex-optimization software~\cite{diamond2016cvxpy,odonoghue2016conic}.

Within the approximations of this convex relaxation, the constraints given by \eqref{eq:constraints} define the joint space of interaction matrices, $\bm{\epsilon}$, and chemical potential vectors, $\vec\mu$, for which bulk phase coexistence can be established among the target condensed phases and a dilute phase. 
We pick out a unique solution from within the solution space by minimizing the nuclear norm, $\|\bm\epsilon\|_*$, and then identify the smallest number of molecular features, $r$, with which we can satisfy the bound given in \eqref{eq:singval-bound} to obtain the low-rank approximation $\bm{\epsilon_r}$.
By contrast, the objective function used in Ref.~\cite{chen2023solver} tends to select $\bm\epsilon$ with larger singular values; as a result, solutions obtained using this alternative objective function tend to require a larger number of molecular features to satisfy the bound given in \eqref{eq:singval-bound}.

\subsection{Step 2: Factorization into molecular features.}
\label{sec:step2}

\subsubsection{Overview of nonnegative matrix factorization for feature-wise interactions}

We next find a molecular realization of the pairwise interactions by factorizing the low-rank interaction matrix, $\bm{\epsilon_r}$.
The rank of the feature interaction matrix $\bm{u}$ must be exactly equal to $r$, so that exactly $r$ features are required for each row vector of the molecular feature matrix $\bm{W}$.
An optimal factorization is obtained by minimizing the Frobenius norm of the reconstruction error,
\begin{equation}
    \label{eq:objective}
    \|\Delta^{\text{recon}}\bm\epsilon\|_F \equiv \left\|\bm{\epsilon_r} - \bm{W} \bm{u} \bm{W}^\top\right\|_F.
\end{equation}
Up until this point, the design approach has been agnostic to the details of the molecular features.
However, the optimal factorization given by \eqref{eq:objective} must typically satisfy additional constraints.
Most importantly, a physically interpretable molecular feature matrix $\bm{W}$ typically has nonnegative entries, since each matrix element indicates the existence and magnitude of a distinct feature within a particular molecule.
In this case, nonnegative matrix factorization (NMF)~\cite{yang2012qnmf} must be used in place of eigenvalue decomposition when solving \eqref{eq:objective}.
Further constraints may also be required depending on the specific requirements of the molecular system and the nature of the molecular features.
For example, if each molecular feature represents the count of a monomer type in a heteropolymer, then the elements of $\bm{W}$ must be nonnegative integers.
Similarly, the feature interaction matrix $\bm{u}$ may either be fixed, as in the case of amino-acid interactions, or designable, as in the case of nucleic acid motifs~\cite{santalucia1998unified}.
Regardless of these system-specific constraints, the reconstruction error must obey $\|\bm{\Delta^{\text{recon}}\epsilon}\|_F \lesssim \min_\alpha \eta\lambda_1^{(\alpha)}$ in order for the designed molecular mixture to form the prescribed condensed phases.

In practice, we factorize the low-rank approximation of the component-wise interaction matrix, $\bm{\epsilon_r}$, by specifying the loss function
\begin{equation}
  \label{eq:L}
  \mathcal{L} \equiv \|\Delta^{\text{recon}}\epsilon\|_F^2 = \| \bm{\epsilon_r} - \bm{W} \bm{u} \bm{W}^\top\|^2_F,
\end{equation}
which is nonlinear with respect to the independent variables $\bm W$ and $\bm u$.
In the following sections, we consider two possible scenarios, in which the molecular feature matrix, $\bm{W}$, can either take nonnegative real values or nonnegative integers.

\subsubsection{NMF without integer constraints}

When there are no integer constraints on the molecular feature matrix $\bm{W}$, \eqref{eq:L} is biconvex with respect to either $\bm W$ or $\bm u$.
The NMF problem can therefore be solved by iteratively alternating optimization of $\bm{W}$ and $\bm{u}$, with guaranteed convergence to a local optimum~\cite{yang2012qnmf}.
Here we consider a scenario in which $\bm W$ takes positive real values (representing, e.g., the fraction of the surface area of a colloidal particle covered by ``patches'' of certain types~\cite{glotzer2007anisotropy}) and $\bm u$ takes negative real values (representing, e.g., the attractive interactions between patches).
To ensure nonnegativity of $\bm W$ and nonpositivity of $\bm u$, we apply a multiplicative update scheme, where $\bm{W}$ and $-\bm{u}$ are multiplied by the ratio of the negative and positive contributions to the corresponding gradient of the loss function.
Specifically, we write the gradient of the loss function $\mathcal{L}$, \eqref{eq:L}, with respect to a matrix $X$ as $\nabla X$.
Then, since all the matrix elements $W_{ij}$ and $-{u}_{ab}$ are positive, we can separate the terms that appear when differentiating \eqref{eq:L} into positive, $\nabla X^+$, and negative, $\nabla X^-$, contributions, such that $\nabla X = \nabla X^+ - \nabla X ^-$.
We then let
\begin{align}
W_{ia} &\leftarrow W_{ia} \left( \frac{\nabla W^-_{ia}}{\nabla W ^+_{ia}} \right)^\gamma, \\
-u_{ab} &\leftarrow -u_{ab} \left( \frac{\nabla (-u)^-_{ab}}{\nabla (-u)^+_{ab}} \right)^\gamma.
\end{align}
This scheme is equivalent to gradient descent with a step size proportional to the ratio $(\nabla^- / \nabla^+)^\gamma$, where $\gamma$ is an adjustable hyperparameter~\cite{lee2000algorithms}.
In the absence of any additional constraints, differentiating \eqref{eq:L} yields the update rules
\begin{align}
  \label{eq:W-update}
  W_{ia} &\leftarrow W_{ia} \bigg(\frac{(\bm\epsilon \bm{W} \bm{u}^\top + \bm\epsilon^\top \bm{W} \bm{u})_{ia}}{(\bm{W} \bm{u} \bm{W}^T \bm{W} \bm{u}^\top + \bm{W} \bm{u}^\top \bm{W}^\top \bm{W} \bm{u})_{ia}} \bigg)^\gamma, \\
  -u_{ab} &\leftarrow -u_{ab} \bigg(\frac{(\bm{W}^\top \bm\epsilon \bm{W})_{ab}}{(\bm{W}^\top \bm{W} \bm{u} \bm{W}^\top \bm{W})_{ab}} \bigg)^\gamma, 
  \label{eq:u-update}
\end{align}
where we choose $\gamma = 1/4$ to ensure stable convergence~\cite{yang2012qnmf}.
We can also include additional constraints, such as minimizing the variance of all monomer--monomer interactions or the variance of homotypic monomer--monomer interactions.
These constraints can be easily implemented by adding the corresponding positive and negative parts of the constraints to the gradient, and then following the update scheme given above.

\subsubsection{NMF with integer constraints}
\label{sec:mip}

Imposing integer constraints on $\bm{W}$ makes the NMF problem no longer convex in $\bm{W}$.
Thus, in cases such as the heteropolymer design problem considered in the main text, where the molecular feature matrix must have integer entries (see also SI~\secref{sec:FH}), optimizing $\bm{W}$ becomes an NP-hard combinatorial optimization problem.
For short polymer chains with $L_{\text{max}} \lesssim 10$ and a prescribed monomer--monomer interaction matrix $\bm u$, it is feasible to find the globally optimal $\bm{W}$ matrix by brute force.
However, brute-force search is infeasible for designing long polymer chains. 
Instead, we follow an approach in which we sample $\bm{W}$ matrices from a probability distribution, and then select candidate $\bm{W}$ matrices from the tail of the reconstruction-error distribution. 
Suppose that every row of the $\bm{W}$ matrix is drawn from a multinomial distribution $\vec W_i \sim \mathcal{M}_r(L_{\max};\vec{p}_i)$ for $i = 1, \ldots, N$, which gives the probability of any particular combination of counts for various feature types $a = 1, \ldots, r$.
The probability vector $\vec p_i$, which is normalized such that $\sum_{a=1}^r p_{ia} = 1$, parameterizes the multinomial distribution.
In the heteropolymer design scenario, the number of categorical variables $r=\text{rank}(\bm\epsilon_r)$ denotes the number of monomer types, $L_{\max}$ is equal to the sum of the counts in $\vec W_i$, and $p_{ia}$ represents the probability of adding a monomer of type $a$ to chain $i$.

We are interested in finding an integer $\bm W$ solution in which the reconstruction error is smaller than a prescribed threshold.
If such $\bm W$ matrices are rare, then we can carry out the search using a cross-entropy (CE) optimization approach~\cite{deboer2005ce}.
This algorithm proceeds as follows:
\begin{enumerate}
  \item Choose an initial parameter matrix $\bm{p}$, which contains all the probability vectors $\vec{p}_i$. Choose the total number of samples to be generated ($N_{\text{samples}} = 10000$) and the number of samples needed for parameter inference ($N_{\text{top}} = 10$).  Let the initial iteration number be $t = 1$. 
  \item To generate a candidate solution pair $(\bm{W}, \bm{u})$, sample each row $\vec W_1, ..., \vec W_N \sim_{\text{iid}} \mathcal{M}_r(L_{\text{max}}, \vec{p}_i)$.  Then, given this candidate $\bm{W}$ matrix, use gradient descent to optimize for a nonpositive $\bm{u}$ matrix by iteratively applying the multiplicative update given in \eqref{eq:u-update} until convergence, defined as the point where the maximum difference between updates of any element of $\bm{u}$ is less than $10^{-5}$.
  \item Repeat step 2 to generate $N_{\text{samples}}$ samples and calculate the loss, \eqref{eq:L}, for each sample.
  \item Update $\bm{p}^{t+1} \leftarrow{{\bm{p}}^{t}}$ according to a maximum likelihood estimate.  Specifically, determine the probability vectors $\vec{p}_{i}^{\;t+1}$ by computing the normalized average frequencies of all monomer types observed in chain type $i$ in the $N_{\text{top}}$ candidate solutions with the smallest values of the loss. 
  \item Iterate steps 2, 3, and 4 until the maximum difference between updates of any element of the parameter matrix $\bm{p}$ is less than 0.005 or the maximal number of iterations (chosen here to be 100) is exceeded. The $(\bm{W}, \bm{u})$ pair with the minimal loss is considered to be a solution if the reconstruction error is below the threshold, \eqref{eq:tolerance}.
\end{enumerate}
We use uniform random initialization of the probability vectors to obtain all the design solutions presented in this paper. 
Because permutation of the rows of $\bm W$ could potentially lead to degenerate solutions during the search, we sort the rows with respect to descending monomer frequencies for every candidate $\bm W$ matrix.
Furthermore, to make the optimization problem easier, we relax the constraint on the degree of polymerization by allowing the sequence length to be shorter than $L_{\text{max}}$.
To this end, we allow $\sum_{a=1}^r p_{ia} \leq 1$, which is realized in practice by adding a dummy column to the probability matrix $\bm{p}$. 
This enhanced-sampling method allows us to efficiently search for NMF solutions for arbitrary long polymer chains.

\section{Multicomponent polymer model and simulation methods}
\label{sec:heteropolymer}

In the main text, we demonstrate the utility of our theory by applying it to a simulation model of heteropolymers in implicit solvent.
In the following sections, we provide a detailed description of the simulation model and then describe how the interaction parameters are determined from the two-step optimization approach described above.

\subsection{Lennard-Jones heteropolymer model}

In our simulations, we consider a simple model of heteropolymers in implicit solvent.
Nonbonded monomers of types $a,b = 1, \ldots, r$ interact via the Lennard-Jones (LJ) pair potential~\cite{lennardjones1924potential},
\begin{equation}
  \label{eq:LJ}
  U^{\text{LJ}}_{ab}(r) = 4 w^{\text{LJ}}_{ab}\left[(d/r)^{12} - (d/r)^6\right] + U_{\text{cut},ab},
\end{equation}
where $w^{\text{LJ}}_{ab} < 0$ represents the interaction strength (i.e., well depth), $d$ is the monomer diameter, and the cutoff distance is $3d$.
The potential is shifted to zero at the cutoff distance by setting $U_{\text{cut},ab} = -4 w^{\text{LJ}}_{ab}[(1/3)^{12} - (1/3)^6]$.
Bonded monomers interact via finite-extensible nonlinear elastic (FENE) bonds~\cite{kremer1990polymer}, which comprise a nonlinear attractive term and a repulsive LJ term.
For the FENE bonds, we use the parameters $(100, 4, 1, 1)$, corresponding to the coefficient of the attractive term, the maximum length of the bond in units of $d$, the well depth for the LJ potential between bonded monomers, and the diameter of the particle in units of $d$.
All simulations are performed using the LAMMPS molecular dynamics package~\cite{plimpton1995fast}.

\subsection{Inverse design of heteropolymers}

\subsubsection{Determining the minimum required number of molecular features}
\label{sec:FH}

The first step in the inverse-design process is optimizing the polymer--polymer interactions.
For this purpose, we utilize the mean-field Flory--Huggins (FH) model to apply our theory to the design of model heteropolymer solutions.
The Helmholtz free-energy density, $f$; chemical potential, $\vec\mu = \partial f / \partial \vec\phi$; osmotic pressure, $P$; and Hessian matrix, $\partial^2 f / \partial\vec\phi^2 = \partial\vec\mu(\vec\phi) / \partial\vec\phi$, in the multicomponent Flory--Huggins model are
\begin{subequations}
  \label{eq:FH}
  \begin{align}
    f &= \sum_{i=1}^N \frac{\phi_i}{L_i} \log \phi_i + (1 - \phiT) \log (1 - \phiT) + \frac{1}{2} \sum_{i=1}^N\sum_{j=1}^N \epsilon_{ij} \phi_i \phi_j \\
    \mu_i &= \frac{1}{L_i} \log \phi_i - \log (1 - \phiT) - \left(1 - \frac{1}{L_i}\right) + \sum_{j=1}^N \epsilon_{ij} \phi_j \\
    P &= -\log (1 - \phiT) + \sum_{i=1}^N \frac{\phi_i}{L_i} - \phiT + \frac{1}{2} \sum_{i=1}^N\sum_{j=1}^N \epsilon_{ij} \phi_i \phi_j \\
    \frac{\partial \mu_i}{\partial \phi_j} &= \frac{\delta_{ij}}{L_i\phi_i} + \frac{1}{1 - \phiT} + \epsilon_{ij},
      \label{eq:dmudphi}
  \end{align}
\end{subequations}
respectively, where $L_i$ is the degree of polymerization of polymeric species $i$.
As noted in SI~\secref{sec:pairwise}, the pairwise interaction matrix $\bm\epsilon$ accounts for the interactions between macromolecules arising from the longer ranged portion of the LJ potential, while the $\phiT$-dependent terms provide an approximate model for the excess chemical potentials of a heteropolymer reference system with soft-core repulsive interactions only.
When applying our theory and design approach to multicomponent polymer solutions, we use these expressions in the convex relaxation, \eqref{eq:constraints}.

\begin{figure}  
  \includegraphics[width=\textwidth]{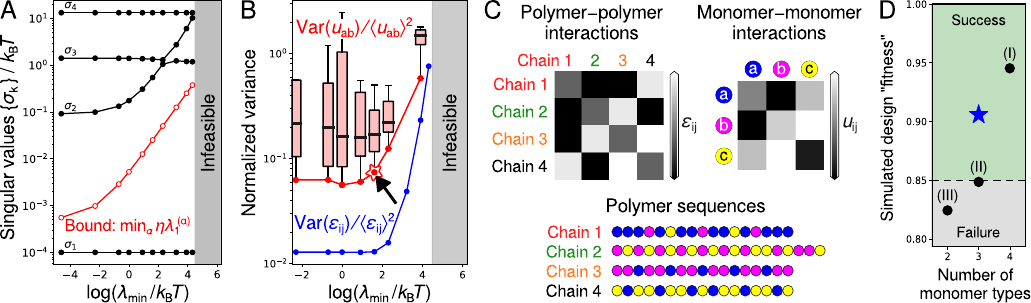}
  \caption{\textbf{Optimization of minimal-complexity interaction matrices.}
    (A)~Inverse design of the 4-component multiphase system illustrated in Fig.~1B in the main text.  The singular values of the optimized interaction matrix $\bm{\epsilon}$, obtained via nuclear-norm minimization, as a function of the landscape stability criterion $\lambda_{\text{min}}$.  The singular values tend to increase as the condensed phases become more stable.  Comparing the singular values with the bound, \eqref{eq:singval-bound}, where $\eta$ is chosen to be equal to $0.1$, indicates that the minimum interaction matrix rank is $r = 3$.
    (B)~The normalized pairwise interaction variance, $\text{Var}(\epsilon_{ij}) / \langle \epsilon_{ij} \rangle^2$, tends to increase with $\lambda_{\text{min}}$.  Box plots show the distribution of normalized monomer--monomer interaction variances obtained by nonnegative matrix factorization (NMF) of the optimized rank-3 polymer--polymer interaction matrices found in \textbf{A}, while treating the distinct monomer types as the molecular features.  All NMF solutions satisfy $\|\bm{\Delta^{\text{recon}}\epsilon}\|_F  / \min_\alpha\lambda_1^{(\alpha)} \le 0.2$.  The solid red line indicates the Pareto front, representing the NMF solutions that simultaneously minimize $\text{Var}(u_{ab})/\langle u_{ab} \rangle^2$ and maximize $\min_\alpha \lambda_1^{(\alpha)}$.
    (C)~The design solution corresponding to the starred location on the Pareto front (see arrow) in \textbf{B}.  The polymer sequences consist of exactly $r=3$ monomer types.  This design and its equilibrium phase behavior are shown in Fig.~2 of the main text.
    (D)~Comparison of the ``fitness'' metric (see SI~\secref{sec:fitness}) for the heteropolymer design shown in \textbf{C} (blue star) and three alternative designs (black circles): (I) a 4-monomer-type design chosen to maximize the condensed-phase stabilities, (II) a 3-monomer-type design constructed by applying the EYM theorem to an interaction matrix that was \textit{not} obtained via nuclear-norm minimization, and (III) a two-monomer-type design constructed by applying NMF with $r=2$ (see SI~\figref{sfig:N4-data}).  [Note that only points (I) and (III) are shown in Fig.~2 of the main text for clarity.]  Designs are classified as successes or failures based on a fitness threshold of 0.85 (see SI~\secref{sec:fitness}).
    \label{sfig:N4-optimization}}
\end{figure}

As an illustrative example for the discussion that follows, we consider the 4-component, 3-condensed-phase mixture depicted in Fig.~1B in the main text, assuming a maximum degree of polymerization of $L_{\text{max}} = 20$.
By solving the minimum-nuclear-norm convex optimization problem as a function of the minimum thermodynamic stability criterion, $\lambda_{\text{min}}$ (SI~\secref{sec:step1}), we obtain the singular values of the optimized pair interaction matrix, $\bm\epsilon$, for this design problem (SI~\figref{sfig:N4-optimization}A).
The nuclear norm of the optimized $\bm\epsilon$ tends to increase with $\lambda_{\text{min}}$, up to the point at which the design problem becomes infeasible, since imposing a stricter constraint on the free-energy landscape tends to shrink the interaction-matrix solution space.
Accordingly, the singular values of $\bm\epsilon$ tend to increase with $\lambda_{\text{min}}$.
We then compare the singular values with the right-hand side of \eqref{eq:singval-bound}, assuming a composition tolerance of $\eta = 20\%$.
Since one singular value lies below the bound in SI~\figref{sfig:N4-optimization}A, \eqref{eq:singval-bound} predicts that one singular value can be eliminated while still satisfying the design problem.
Thus, only $r=3$ molecular features are required for the construction of a heteropolymer mixture that phase-separates into the nontrivial phases shown in Fig.~1B in the main text.

\subsubsection{Factorization into monomer--monomer interactions}
\label{sec:LJ-factorization}

We now assume that the molecular features represent distinct monomer types in the LJ heteropolymer model in order to factorize $\bm{\epsilon_r}$ into a molecular feature matrix, $\bm{W}$, and a monomer--monomer interaction matrix, $\bm{u^{\text{LJ}}}$.
With this choice, each row of the $N \times r$ $\bm{W}$ matrix represents a count encoding of the number of occurrences of each monomer type, $a = 1,\ldots,r$, within each heteropolymer type, $i=1,\ldots,N$.
All elements of $\bm{W}$ are therefore nonnegative integers, and the row sums of $\bm{W}$ are bounded by the maximum degree of polymerization $L_{\text{max}}$.
Meanwhile, $\bm{u^{\text{LJ}}}$ is a nonpositive $r \times r$ interaction matrix determined by the attractive portion of the monomer--monomer LJ pair potential.
Here, assuming that the molecular features represent distinct monomer types is equivalent to a pairwise-additive approximation for the polymer--polymer interactions, such that $\epsilon_{ij} = \sum_{a,b} W_{ia}u^{\text{LJ}}_{ab}W_{jb}$.
This factorization of $\bm\epsilon$ follows directly from the Flory--Huggins model, since the pairwise portion of the free-energy density, $(1/2)\sum_{i,j}\epsilon_{ij}\phi_i\phi_j$, represents the mean-field interaction between all pairs of monomers in a polymer solution~\cite{colby2003polymer}.
More specifically, we can write this sum in the factorized form $(1/2)\sum_{i,j}\sum_{a,b}u_{ab}W_{ia}W_{jb}\phi_i\phi_j = (1/2)\sum_{a,b} u_{ab}\tilde\phi_a\tilde\phi_b$, where $\tilde\phi_a \equiv \sum_iW_{ia}\phi_i$ is the volume fraction occupied by monomers of type $a$.
We emphasize that this approximation is not required by our theory, since the molecular features could represent interactions between units larger than individual monomers, such as short motifs in nucleic acid sequences.
However, we choose to make this assumption here to provide a transparent proof of principle of our theory.
We further note that this approximation establishes an upper bound on the required number of distinct monomer types, since the number of molecular features obtained from an eigendecomposition of $\bm\epsilon$ cannot be less than the number of distinct monomer types.
In other words, while we can use this approximation to rigorously test our theory, the bound on the required number of monomer types could potentially be improved by relaxing this assumption.

Given this choice of molecular features, we solve \eqref{eq:objective} subject to the optimized low-rank interaction matrix, $\bm{\epsilon_r}$, and the constraints on $\bm{W}$ and $\bm{u}$.
Following the algorithm described in SI~\secref{sec:mip}, we generate an ensemble of $\bm{W}$ matrices and iteratively solve for the interaction matrix $\bm{u}$ to ensure nonpositivity; we then utilize enhanced sampling to probe the tail of the reconstruction-error distribution to find the $\bm{W}$ matrix with the smallest $\|\bm{\Delta^{\text{recon}}\epsilon}\|_F$.
Representative results of this stochastic algorithm are shown in SI~\figref{sfig:N4-optimization}B.
We find that the relative variance of the optimized monomer--monomer interactions, $\text{Var}(u_{ab})/\langle u_{ab} \rangle^2$, tends to increase with the minimum stability constraint $\lambda_{\text{min}}$.
This trend reflects the behavior of the polymer--polymer interaction matrix, $\bm{\epsilon_r}$, that is being factorized, $\text{Var}(\epsilon_{ij})/\langle \epsilon_{ij} \rangle^2$, and suggests a fundamental trade-off: Designing highly stable condensates, which are least sensitive to interaction-matrix perturbations according to \eqref{eq:response}, requires highly dissimilar monomer--monomer interactions.

Finally, we use an optimized monomer-composition matrix, $\bm{W}$, and monomer--monomer interaction matrix, $\bm{u^{\text{LJ}}}$, to construct a set of heteropolymer sequences for simulation.
Because the pairwise-additive approximation relating $\epsilon_{ij}$ and $u_{ab}^{\text{LJ}}$ is most accurate when $\text{Var}(u_{ab})/\langle u_{ab} \rangle^2$ is small, we select monomer designs from the Pareto front shown in SI~\figref{sfig:N4-optimization}B.
We then compose the polymer sequences from the monomer compositions by interleaving the different monomer types to minimize the blockiness of each sequence, as we describe in the next section.
An example outcome of this algorithm, which we will directly test via molecular simulation, is shown in SI~\figref{sfig:N4-optimization}C.

\subsubsection{Constructing LJ interaction parameters and heteropolymer sequences}

To perform simulation tests of our heteropolymer mixture designs, we need to construct LJ interaction matrices, $\{w^{\text{LJ}}_{ab}\}$, and polymer sequences using the optimal $\bm{u^{\text{LJ}}}$ and $\bm{W}$ matrices obtained from the previous section.
We describe each of these steps in turn.

First, to convert between the mean-field monomer--monomer interaction coefficients, $\{u^{\text{LJ}}_{ab}\}$, and the well depths for the LJ interactions, $\{w^{\text{LJ}}_{ab}\}$, we perform a nonlinear mapping by matching the longer-ranged contributions (i.e., excluding the soft-core repulsions) to the monomer--monomer second viral coefficients.
Specifically, we relate the attractive portion of the LJ second virial coefficient, for $r > d$, to $u^{\text{LJ}}_{ab}$,
\begin{equation}
  \label{eq:u-int}
  \frac{u_{ab}^{\text{LJ}}}{|\bar u^{\text{LJ}}|} = \frac{2\pi}{d^3} \int_d^{3d} dr\,r^2 \left\{ 1 - \exp[-U_{ab}^{\text{LJ}}(r) / k_{\text{B}}T] \right\},
\end{equation}
given a fixed temperature $T$.
For simplicity, we work at a standard temperature, such that $k_{\text{B}}T = 1$ in our simulations.
The user-defined scale factor, $|\bar u^{\text{LJ}}|$, controls the mean LJ interaction strengths used in the simulations; this scale factor must be introduced empirically (as opposed to being determined from the mean-field model) because the critical points of the mean-field FH model and the LJ heteropolymer simulations differ.
In practice, we determine the LJ coefficients $\{w^{\text{LJ}}_{ab}\}$ from a designed monomer--monomer interaction matrix, $\bm{u^{\text{LJ}}}$, by choosing an appropriate value of $|\bar u^{\text{LJ}}|$ (determined from simulations of homomeric LJ heteropolymers with chain length $L_{\text{max}}$) and inverting \eqref{eq:u-int}.

Second, to design the polymer sequences for polymer species $i = 1,\ldots,N$, we utilize the count encodings in each row, $\vec W_i$, of the optimized molecular feature matrix.
We aim to construct sequences that are minimally ``blocky'', since the pairwise-additive assumption used in our heteropolymer design approach (see SI~\secref{sec:LJ-factorization}) does not consider sequence-patterning effects.
We therefore aim to construct sequences in which the monomers of each type are homogeneously distributed throughout each polymer chain.
Here we use a deterministic heuristic for interleaving different monomer types.  
To generate the sequence design for chain $i$, we first sort the monomer frequency counts $\{W_{ia}\}$ in descending order and associate the $n$th monomer of type $a$ with the fractional number $n/(1+W_{ia})$, where $n$ goes from $1$ to $W_{ia}$.
We then read off the order of the monomer types in the sequence by going through the array of fractional numbers in ascending order.

Following these procedures, we obtain the LJ coefficients used for simulating the heteropolymer design shown in Fig.~2 in the main text,
\begin{equation*}
  \bm{w^{\text{LJ}}} = 
  \begin{bmatrix}
    0.520 & 0.863 & 0.380 \\
    0.863 & 0.345 & 0.238 \\
    0.380 & 0.238 & 0.788      
   \end{bmatrix}.
\end{equation*}
These LJ coefficients result in the $\bm{u^{\text{LJ}}}$ matrix shown in SI~\figref{sfig:N4-optimization}C and in Fig.~2A of the main text.
The corresponding sequences are also shown in SI~\figref{sfig:N4-optimization}C and in Fig.~2A of the main text.

\subsection{Direct-coexistence simulations}

\subsubsection{Initialization of multiphase simulations}

To prepare condensed phases for molecular dynamics simulations, we first perform constant-temperature-and-pressure (NPT) simulations for each individual target phase.
We use $N_{\text{tot}} = 432$ chains in a cubic box with periodic boundary conditions, where chain types are assigned according to the target-phase composition.
Performing NPT simulations at zero pressure allows us to estimate the average equilibrium total number density in a condensed phase, $\rho_{\text{eq}}$.
We then prepare the condensed phases for use in direct-coexistence simulations by deforming the simulation box.
In this step, we maintain periodic boundary conditions in both the x and y directions while imposing hard wall constraints and open boundary conditions in the z direction.
The final dimensions, consistent with the target density $\rho_{\text{eq}}$ determined from the NPT simulations, are $l_z = 40 d$ and $l_x = l_y = 20 d$ for simulations of designs with $L_{\text{max}} = 20$, and $l_z = 40 d$ and $l_x = l_y = 24.5 d$ for simulations of designs with $L_{\text{max}} = 30$.
Finally, the initial condition for a direct-coexistence simulation of a pair of condensed phases is constructed by stitching together two condensed phases and a dilute gas phase along the z axis of the simulation box.
The region of the simulation box corresponding to the dilute phase is initialized with 6 chains of each type at a total number density of $\rho_{\text{dilute}} = 1.50 \times 10^{-3}$, $2.25 \times 10^{-3}$, or $1.50 \times 10^{-3}$ for the designs shown in Fig.~2B, Fig.~4A, and Fig.~4C in the main text, respectively.

\subsubsection{Production simulations}

In our production runs, we perform constant-temperature-and-volume (NVT) direct-coexistence simulations.
The overall dimensions of the simulation box, including $\alpha$, $\beta$, and dilute phases, are $l = 20 d \times 20 d \times 120 d$ for simulations of designs with $L_{\text{max}} = 20$, and $l = 24.5 d \times 24.5 d \times 120 d$ for simulations of designs with $L_{\text{max}} = 30$.
Each direct-coexistence simulation contains a total of $N_{\text{tot}} = 888$ chains for the four-component case and $N_{\text{tot}} = 900$ chains for the six-component cases.
Our molecular dynamics simulations utilize the Nose--Hoover thermostat and a timestep of $5\times 10^{-3}$ in LJ units.

\begin{figure} 
 \includegraphics[width=\columnwidth]{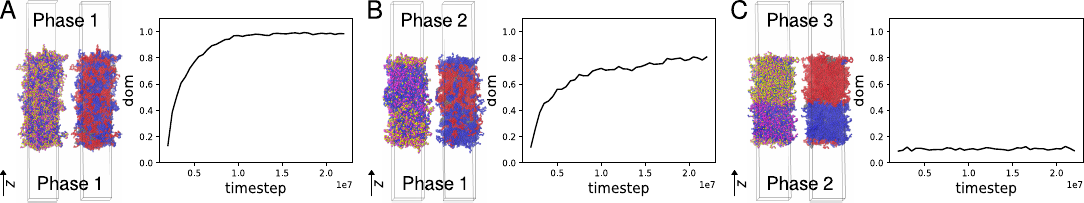}
  \caption{\textbf{The degree of mixing as an indicator of equilibration.}
    For the example design solution shown in Fig.~2 in the main text, we compare the evolution of the degree of mixing (d.o.m.) between (A)~an $\alpha=\beta$ control simulation and (B--C)~two $\alpha\ne\beta$ direct-coexistence simulations.  The left and right snapshots in each panel are colored according to the monomer types and the initial phase labels (either red or blue) at the end of the simulation trajectory.  From \textbf{A}, we determine that the mixing timescale at this temperature and
    mean LJ interaction strength is approximately $10^7$ timesteps ($5\times 10^{4}\tau$).  The d.o.m.\ is significantly greater than zero in \textbf{B} because phases 1 and 2 share an enriched component.  Interfacial effects, both at the $\alpha$--$\beta$ interface and at the condensed--dilute interfaces, also tend to increase the d.o.m.\ in both \textbf{B} and \textbf{C}.
  \label{sfig:dom}}
\end{figure}

To verify equilibration, we calculate the degree of mixing (d.o.m.) over the region of the simulation box that is occupied by condensed phases (SI~\figref{sfig:dom}).
The d.o.m.\ is defined to be
\begin{equation}
  \text{d.o.m.} \equiv 1 - \frac{1}{2} \int_{\phiT(z)>\nicefrac{1}{2}} dz\, \frac{\left[p_{\text{init}}^{(\beta)}(z) - p_{\text{init}}^{(\alpha)}(z)\right]^2}{p_{\text{init}}^{(\alpha)}(z) + p_{\text{init}}^{(\beta)}(z)},
\end{equation} 
where $p_{\text{init}}^{(\alpha)}(z)$ and $p_{\text{init}}^{(\beta)}(z)$ represent the probability of finding a chain that was initialized in the $\alpha$ or $\beta$ condensed phase, respectively, at the position $z$ along the simulation box.
These probabilities are normalized such that $\int_{\phiT(z)>\nicefrac{1}{2}} dz\, p_{\text{init}}^{(\alpha)} = 1$.
If the chains completely mix, such that there is no correlation between the location of a particular chain and the condensed phase in which it was initially placed, then the degree of mixing tends to one.
If the chains remain in their original phases, then the degree of mixing is close to zero.
In situations where one or more species are shared between a pair of condensed phases, then the degree of mixing is expected to plateau at a value between zero and one.
In practice, we use the degree of mixing to determine the timescale over which equilibration takes place in $\alpha = \beta$ control simulations (SI~\figref{sfig:dom}A).
We also verify that the degree of mixing for $\alpha\neq\beta$ direct-coexistence simulations reaches a plateau value in the course of a simulation trajectory that is at least twice as long as the control-simulation mixing timescale (SI~\figref{sfig:dom}B--C).

\subsection{Defining the heteropolymer design ``fitness'' metric}
\label{sec:fitness}

To quantify the results of our equilibrium coexistence simulations and to compare alternative polymer designs, we introduce an intuitive fitness metric based on the target order parameter profiles,
\begin{equation}
  \label{eq:fitness}
  \text{fitness} \equiv \frac{\int_{\phiT(z)>\nicefrac{1}{2}} dz\, \max_\alpha\! \left[q^{(\alpha)}(\vec\phi(z))\right]}{\int_{\phiT(z)>\nicefrac{1}{2}} dz},
\end{equation}
where the order parameter for the $\alpha$ target phase is defined to be
\begin{equation}
  \label{eq:order-parameter}
  q^{(\alpha)}(\vec\phi) \equiv \frac{ \vec\phi \cdot  \vec\phi^{(\alpha)} }{ ||\vec\phi|| \, ||\vec\phi^{(\alpha)}|| }.
\end{equation}
The fitness metric is equal to one if the bulk molecular concentrations in the condensed phases are precisely equal to the target molecular concentrations and the interfaces are perfectly sharp.
However, in direct-coexistence simulations with $\alpha$, $\beta$, and dilute phases, a valid design will lead to (at least) three finite-width interfaces.
A lower bound on the fitness of a valid design can be estimated by assuming that the integrand in \eqref{eq:fitness} is zero in the interfacial regions and that there are three interfacial regions, each of which comprises at most $5\%$ of the total region in which $\phiT \ge \nicefrac{1}{2}$ (based on the typical interfacial width in the control simulations shown in SI~\figref{sfig:dom}A).
In this way, we estimate that a valid design should have a fitness score of at least $0.85$.
We use this value as the threshold for determining whether a heteropolymer design is a success or a failure in Figs.\ 2 and 4 in the main text.
We note that this heuristic fitness metric might not work as well as the number of components increases, since the order parameter is based on a Euclidean distance, and it generally becomes harder to distinguish points in this way in high-dimensional spaces.

\begin{figure} 
\includegraphics[width=\columnwidth]{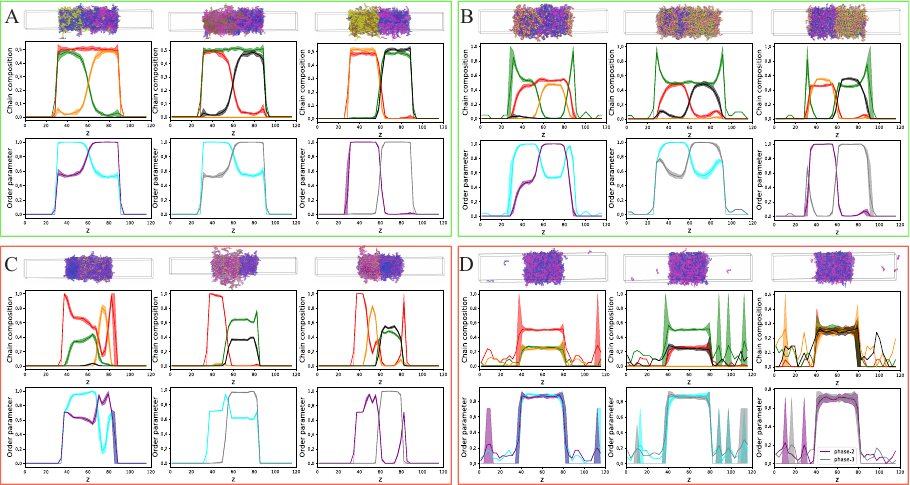}
\caption{\textbf{Simulation results for alternative heteropolymer design strategies}.
  Simulation results are shown for the alternative LJ heteropolymer designs summarized in Fig.~2C of the main text and in SI~\figref{sfig:N4-optimization}D.
  (A)~A full-rank $r = 4$ design.  The interaction matrix is obtained by minimizing the variance of the independent entries of $\bm\epsilon$, as in Ref.~\cite{chen2023solver}.
  (B)~The $r = 3$ design shown in Fig.~2A in the main text.
  (C)~A design obtained by running the NMF solver on the interaction matrix used in \textbf{A}, but enforcing $r = 3$ monomer types.
  (D)~A design obtained by running the NMF solver on the optimized $r=3$ interaction matrix used in \textbf{B}, but enforcing $r = 2$ monomer types.
  The green and red outlines around each panel indicate whether the design is determined to be a success or a failure, respectively, according to the fitness metric, \eqref{eq:fitness}.
  \label{sfig:N4-data}}
\end{figure}

Using this fitness metric, we are able to objectively and consistently determine whether a heteropolymer mixture design results in the target multicomponent phase behavior.
The equilibrated concentration and order-parameter profiles of the alternative designs for the example four-component design problem (see Fig.~1B and Fig.~2 in the main text) are shown in SI~\figref{sfig:N4-data}.
The corresponding fitness values for these four designs are shown in SI~\figref{sfig:N4-optimization}D.

\subsection{Computing effective polymer--polymer interactions in dilute and condensed phases}
\label{sec:effective-interactions}

\subsubsection{Effective polymer--polymer interactions in the dilute limit}

\begin{figure} 
 \includegraphics[width=0.75\columnwidth]{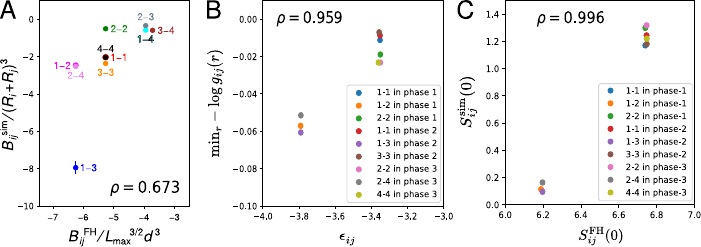}
 \caption{\textbf{Effective polymer--polymer interactions determined from dilute-phase calculations and radial distribution functions (RDFs) in condensed phases.} 
   (A)~Correlation between the second virial coefficients obtained from simulations and predicted by the Flory--Huggins (FH) model.  Labels indicate chain pairs.  Simulated coefficients are normalized by $(R_i+R_j)^3$, where $R_i$ is the radius of gyration of chain $i$ in isolation, while predicted coefficients are normalized by the pervaded volume of an ideal polymer with chain length $L_{\text{max}}$.
   (B)~Correlations between the effective well depths inferred from the RDFs (see Fig.~3B in the main text) and the pairwise interactions, $\epsilon_{ij}$.  The legend indicates pairs of chain types.
   (C)~Correlations between the zero-mode partial structure factors, $S_{ij}(0)$, determined from the RDFs and the mean-field interactions in the Flory--Huggins model.
   In all panels, the Pearson correlation coefficient, $\rho$, is indicated.
   \label{sfig:rdf}}
\end{figure}

We first consider the interactions between polymers in a dilute mixture.
Polymer--polymer interactions in the dilute limit can be quantified by second virial coefficients~\cite{hansen2013theory},
\begin{equation}
  \label{eq:Bijsim}
  B_{ij}^{\text{sim}} = 2\pi \int_0^\infty dr\,r^2 \left\{1 - \exp\left[-w_{ij}(r) / k_{\text{B}}T\right]\right\},
\end{equation}
where $w_{ij}(r)$ is the potential of mean force (PMF) between the centers of mass of two polymers of types $i$ and $j$ at infinite dilution.
In practice, we compute $w_{ij}(r)$ using adaptive biasing force (ABF) simulations~\cite{darve2008adaptive} implemented via the COLVARS package~\cite{henin2009colvars} in LAMMPS~\cite{plimpton1995fast}.
We run constant-temperature-and-volume (NVT) simulations with two chains initialized in a $20d \times 20d \times 20d$ simulation box with shrink-wrapped boundary conditions.
Force statistics are stored in bins of width $0.25d$.
The biasing force is applied once 1000 samples are collected in each bin, after which a final production run is performed for $10^8$ steps with a timestep of $0.005\tau$ (in LJ units).
The second virial coefficient is then obtained by integrating the PMF using \eqref{eq:Bijsim}.
In SI~\figref{sfig:rdf}A, five independent ABF simulations are performed for each $B_{ij}$ calculation to determine the statistical errors, which are shown as error bars.

We then compare the second virial coefficients computed via simulation to those predicted by the Flory--Huggins model,
\begin{equation}
  B_{ij}^{\text{FH}} = \frac{d^3 L_iL_j(1 + \epsilon_{ij}/k_{\text{B}}T)}{2},
\end{equation}
obtained via a power-series expansion of the pressure in \eqref{eq:FH}~\cite{li2023interplay}.
This comparison is shown in SI~\figref{sfig:rdf}A.
We find that $B_{ij}^{\text{sim}}$ and $B_{ij}^{\text{FH}}$ are positively correlated with a Pearson correlation coefficient of 0.67, suggesting non-negligible differences between the dilute-phase polymer--polymer interactions and the predictions of the pairwise-additive approximation.
This discrepancy is expected, since the interactions between pairs of polymer chains in dilute solution are, in general, rarely described accurately by mean-field approximations~\cite{colby2003polymer}.

\subsubsection{Effective polymer--polymer interactions in condensed phases: Excess chemical potential differences}

Turning to the condensed phases, we extract the excess chemical potential differences, $\Delta_{\alpha \beta}\mu_{\text{ex}, i} \equiv \mu^{(\beta)}_{\text{ex},i} - \mu^{(\alpha)}_{\text{ex},i}$, from the equilibrium compositions determined from direct-coexistence simulations of simulated $\alpha$ and $\beta$ bulk phases via~\cite{jacobs2023theory}
\begin{equation}
  \label{eq:muex-diff}
  \Delta_{\alpha \beta}\mu^{\text{sim}}_{\text{ex,}i} = k_{\text{B}} T \log \left( \langle\phi_i^{(\alpha)}\rangle / \langle\phi_i^{(\beta)}\rangle \right).
\end{equation}
We then compare these calculations with the Flory--Huggins prediction,
\begin{equation}
  \Delta_{\alpha \beta}\mu^{\text{FH}}_{\text{ex}, i} = -k_{\text{B}}T \log[(1-\phiT^{(\beta)}) / (1-\phiT^{(\alpha)})] + \sum_j \epsilon_{ij} (\phi_j^{(\beta)} - \phi_j^{(\alpha)}).
\end{equation}
This comparison, shown in Fig.~3A in the main text (where $\Delta_{\alpha \beta}\mu^{\text{FH}}_{\text{ex}, i}$ is labeled $\Delta_{\alpha \beta}\mu^{\text{pw}}_{\text{ex}, i}$) demonstrates a strong correlation ($\rho = 0.953$) between the predicted and measured excess chemical potential differences in the condensed phases.
This result supports our conclusion that the pairwise approximation is best suited to the description of condensed phases.

\subsubsection{Effective polymer--polymer interactions in condensed phases: Radial distribution functions and structure factors}

We can also assess the accuracy of the pairwise-additive approximation by analyzing the microstructure of simulated condensed phases.
For each condensed phase, we run an NPT simulation at zero pressure to compute the radial distribution function (RDF) with respect to the center of mass distance between polymers of various species.
The RDF between chain types $i$ and $j$ is denoted $g_{ij}(r)$.
The results of these simulations are shown in Fig.~3B in the main text; note that these calculations can only be performed with sufficient statistical accuracy for chain types that are enriched in a particular condensed phase.
From the RDFs, we measure the well depths of the effective interactions, $\min_r -\log g_{ij}(r)$, which strongly correlate with the predicted pairwise interactions, $\epsilon_{ij}$ ($\rho = 0.959$; SI~\figref{sfig:rdf}B).
This correspondence provides additional support for our conclusion that the mean-field approximations hold well in the condensed phases.

The mean-field interactions can also be related to the RDFs using linear response theory~\cite{hansen2013theory}.
Specifically, the partial structure factor is related to the Fourier transform of the pair correlation function, $\hat h \equiv \hat g - 1$, via
\begin{equation}
    S_{ij} (\vec{k}) = x_i \delta_{ij} + x_i x_j \rho \hat{h}_{ij}(\vec{k}),
\end{equation}
where $x_i = \phi_i/\phiT$ is the chain composition in a condensed phase, $\rho = N_{\text{tot}}/V$ is the total number density in the condensed phase, and $V$ is the simulated volume of the condensed phase.
It follows that
\begin{equation}
  \begin{aligned}
    \lim_{|\vec{k}| \to 0 }S_{ij} (\vec{k}) 
    &= x_i \delta_{ij} + 2 x_i x_j \rho  \int_0^\infty dr\, r h(r) \times
    \lim_{|\vec{k}| \to 0} \frac{\sin(2\pi r |\vec{k}|)}{|\vec{k}|} \\
    &= x_i \delta_{ij} + 4\pi x_i x_j \rho \int_0^\infty dr\, r^2 h(r).
  \end{aligned}
  \label{eq:structure}
\end{equation} 
Finally, from linear response theory~\cite{hansen2013theory}, we find
\begin{equation}
    S_{ij} (0) = V^{-1} \frac{\partial \langle N_i \rangle}{\partial \beta \mu_j} = k_{\text{B}} T \left(\frac{\partial \mu_j}{\partial \rho_i} \right)^{-1} = k_{\text{B}} T \left(L_i \frac{\partial \mu_j}{\partial \phi_i} \right)^{-1}.
\end{equation}
We calculate the zero mode for number density fluctuations, $S_{ij}(0)$, via the mean-field expression, \eqref{eq:FH}d, and the simulated single-phase RDFs, \eqref{eq:structure}, and find that the results are highly correlated ($\rho = 0.996$; SI~\figref{sfig:rdf}C).
This finding also supports our conclusion that the mean-field approximations hold well in the condensed phases.

\subsubsection{Calculation of overlap parameter}

We compute the overlap parameter, $P$~\cite{colby2003polymer}, from single-phase NPT simulations by estimating the median number of distinct neighboring chains with which every chain in a condensed phase interacts.
A chain is considered to be an interacting neighbor if there is at least one inter-chain pair of monomers that are within the cutoff distance, $3d$, of the monomer--monomer LJ potential.
The relatively large average value of the overlap parameter, $P \approx 13 \pm 2$, is consistent with the observed mean-field behavior of the condensed phases.

\section{Design and simulation data for 6-component heteropolymer designs}

\begin{figure} 
  \includegraphics[width=\columnwidth]{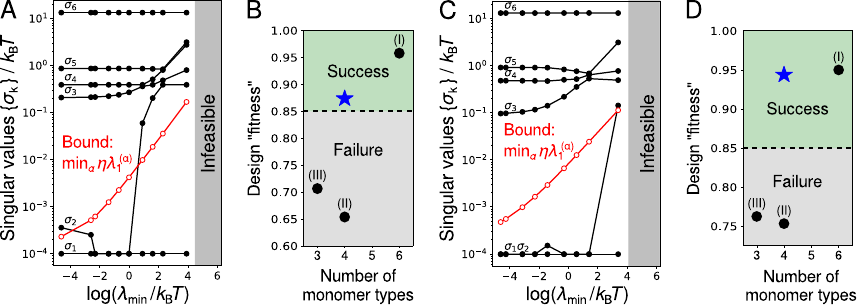}
  \caption{\textbf{Convex optimization results for example six-component condensate design problems.}
    (A)~Convex optimization results for the condensate design problem shown in Fig.~4A in the main text, using $\eta = 0.1$ for the calculation of the bound.
    Data are shown as in SI~\figref{sfig:N4-optimization}A.
    (B)~Comparison of the optimized design shown in \textbf{A} with alternative designs, as in SI~\figref{sfig:N4-optimization}D.  The alternative designs (I) maximize the stability of the condensed phases using six monomer types, (II) apply the EYM theorem without nuclear-norm minimization, and (III) apply NMF with $r=3$ monomer types.
    (C)~Convex optimization results for the condensate design problem shown in Fig.~4C in the main text, using $\eta = 0.1$ for the calculation of the bound.
    (D)~Comparison of the optimized design shown in \textbf{C} with alternative designs, following the same definitions as in~\textbf{B}.
    \label{sfig:N6-singval}}
\end{figure}

In this section, we present the optimization results and resulting simulation data for the problems considered in Figs.\ 4A and 4C in the main text.
The results of convex optimization for both problems are shown in SI~\figref{sfig:N6-singval}A,C using the same format as in SI~\figref{sfig:N4-optimization}A.
For the condensate design problem shown in Fig.~4A, we select the optimal solution using the parameters $\lambda_{\min} = 0.5 k_{\text{B}}T$ and $L_{\max} = 20$.
For the condensate design problem shown in Fig.~4C, we select the optimal solution using the parameters $\lambda_{\min} = 0.075 k_{\text{B}}T$ and $L_{\max} = 30$.

\begin{figure}
  \includegraphics[width=0.85\columnwidth]{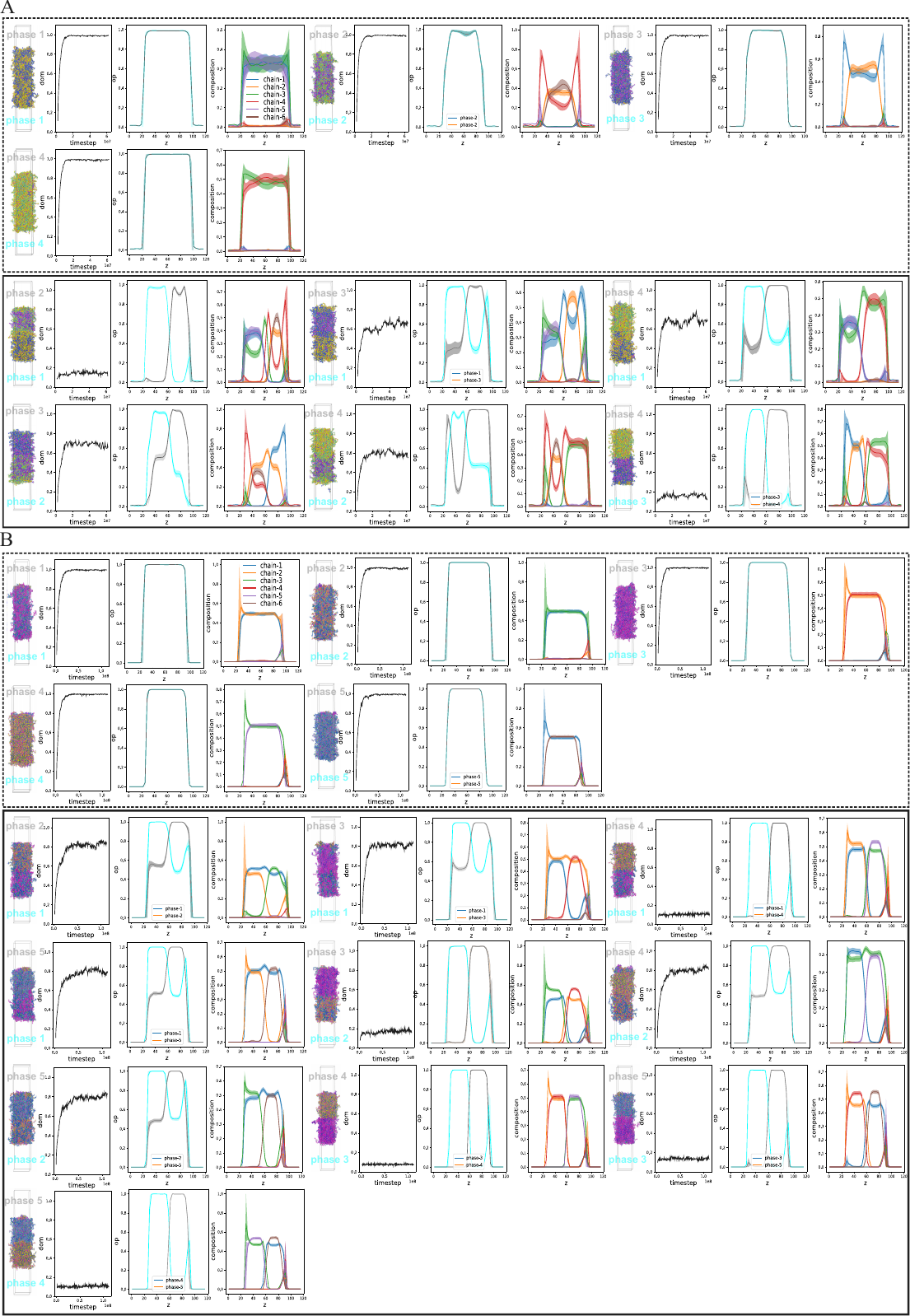}
  \caption{\textbf{Molecular simulation results for optimized LJ heteropolymer solutions to example six-component condensate design problems.}
    Simulation results of all pairs of coexisting phases are shown for (A)~the condensate design problem presented in Fig.~4A of the main text and (B)~Fig.~4C of the main text.
    Results in dashed boxes correspond to control simulations, for which $\alpha=\beta$.
    Results in solid boxes correspond to direct-coexistence simulations of immiscible phases, for which $\alpha \ne \beta$.
  \label{sfig:N6-data}}
\end{figure}

The LJ coefficients used for simulating the heteropolymer design shown in Fig.~4A in the main text are
\begin{equation*}
  \bm{w^{\text{LJ}}} =
    \begin{bmatrix}
    0.630 & 0.559 & 0.430 & 0.264 \\
    0.559 & 0.582 & 0.374 & 0.877 \\
    0.430 & 0.374 & 0.516 & 0.361 \\
    0.264 & 0.877 & 0.361 & 0.663
    \end{bmatrix}.
\end{equation*} 
The LJ coefficients used for simulating the heteropolymer design shown in Fig.~4C in the main text are
\begin{equation*}
  \bm{w^{\text{LJ}}} = 
    \begin{bmatrix}
    0.135 & 0.701 & 0.482 & 0.457 \\
    0.701 & 0.483 & 0.271 & 0.375 \\
    0.482 & 0.271 & 0.951 & 0.525 \\
    0.457 & 0.375 & 0.525 & 0.982
    \end{bmatrix}.
\end{equation*}
The corresponding simulation data for these designs, along with the optimized sequences shown in Figs.\ 4A and 4C in the main text, are shown in SI~\figref{sfig:N6-data}.
The simulation fitness metrics for these designs, as well as for alternative heteropolymer mixture designs that do not follow our two-step optimization approach, are shown in SI~\figref{sfig:N6-singval}B,D and for each pair of phases ($\alpha,\beta$) in Fig.~4 of the main text.

\providecommand{\latin}[1]{#1}
\makeatletter
\providecommand{\doi}
  {\begingroup\let\do\@makeother\dospecials
  \catcode`\{=1 \catcode`\}=2 \doi@aux}
\providecommand{\doi@aux}[1]{\endgroup\texttt{#1}}
\makeatother
\providecommand*\mcitethebibliography{\thebibliography}
\csname @ifundefined\endcsname{endmcitethebibliography}
  {\let\endmcitethebibliography\endthebibliography}{}